\documentclass[aps,onecolumn,superscriptaddress]{revtex4-2}
\usepackage[utf8]{inputenc}
\usepackage[colorlinks=true,linkcolor=blue,citecolor=blue,urlcolor=blue]{hyperref}
\hypersetup{pdfborder={0 0 0}}

\usepackage{graphicx}  
\usepackage{dcolumn}   
\usepackage{bm}        
\usepackage{amssymb,amsmath}   
\usepackage{uri}
\usepackage{float}
\usepackage{geometry}
\usepackage{comment}
\usepackage{fixmath}
\newcommand{\drk}{\delta\rho(\bm{k},t)}
\newcommand{\drak}{\delta\rho_a(\bm{k},t)}
\newcommand{\drbk}{\delta\rho_b(\bm{k},t)}
\newcommand{\omOne}{\bm{\Omega}'}
\newcommand{\omTwo}{\bm{\Omega}^{\prime \prime}}

\newcommand{\dpak}{\delta \bm{p}_a(\bm{k},t)}

\newcommand{\dck}{\delta c(\bm{k},t)}

\newcommand{\dwk}{\delta p_{\parallel}(\bm{k},t)}
\newcommand{\dwak}{\delta p_{\parallel,a}(\bm{k},t)}
\newcommand{\dwbk}{\delta p_{\parallel,b}(\bm{k},t)}

\newcommand{\dperk}{\delta \bm{p}_{\perp}(\bm{k},t)}
\newcommand{\dperak}{\delta \bm{p}_{\perp,a}(\bm{k},t)}

\DeclareMathOperator{\sign}{sign}
\DeclareMathOperator{\Trace}{tr}
\DeclareMathOperator{\Det}{det}

\begin{document}
\title{Nonreciprocal collective dynamics in a mixture of phoretic Janus colloids}

\author{Gennaro Tucci}
\affiliation{Max Planck Institute for Dynamics and Self-Organization (MPIDS), D-37077 Göttingen, Germany}
\author{Ramin Golestanian}
\email{ramin.golestanian@ds.mpg.de}
\affiliation{Max Planck Institute for Dynamics and Self-Organization (MPIDS), D-37077 Göttingen, Germany}
\affiliation{Rudolf Peierls Centre for Theoretical Physics, University of Oxford, Oxford OX1 3PU, United Kingdom}
\author{Suropriya Saha}
\email{suropriya.saha@ds.mpg.de}
\affiliation{Max Planck Institute for Dynamics and Self-Organization (MPIDS), D-37077 Göttingen, Germany}



\begin{abstract}
A multicomponent mixture of Janus colloids with distinct catalytic coats and phoretic mobilities is a promising theoretical system to explore the collective behavior arising from nonreciprocal interactions. An active colloid produces (or consumes) chemicals, self-propels, drifts along chemical gradients, and rotates its intrinsic polarity to align with a gradient. As a result the connection from microscopics to continuum theories through coarse-graining couples densities and polarization fields in unique ways. Focusing on a binary mixture, we show that these couplings render the unpatterned reference state unstable to small perturbations through a variety of instabilities including oscillatory ones which arise on crossing an exceptional point or through a Hopf bifurcation. For fast relaxation of the polar fields, they can be eliminated in favor of the density fields to obtain a microscopic realization of the Nonreciprocal Cahn-Hilliard model for two conserved species with two distinct sources of non-reciprocity, one in the interaction coefficient and the other in the interfacial tension. Our work establishes Janus colloids as a versatile model for a bottom-up approach to both scalar and polar active mixtures. 
\end{abstract}

\maketitle

\section{Introduction}
 As a route to manifesting active matter systems \cite{Gompper2020}, the breaking of action-reaction symmetry in effective interactions, or non-reciprocity, has recently garnered increased attention~\cite{Challenges_PhysRevX.12.010501}. Reciprocity in interactions is synonymous with the existence of an interaction potential. If the concept of an effective free energy cannot be applied, as is very likely to be the case for interactions driven by chemicals~\cite{ThermalActiveColloids,Golestanian2019phoretic}, social interactions~\cite{SocialForceModel}, velocity fields~\cite{Uchida2010_PRL,DavidPNAS,DancingVolvox}, or information transfer~\cite{Ziepke2022,Klapp_Loos_2020,Osat2023} non-reciprocity will inevitably emerge, whether or not it is significant at long timescales~\cite{Dinelli2023}. 

In a system of particles without an intrinsic polarity, non-reciprocity is apparent only in active mixtures which allow breaking Newton's third law in pairwise interactions. For example, activity manifests in the formation of novel bound states in collections of uniformly coated active colloids~\cite{SotoPhysRevLett.112.068301, Soto2PhysRevE.91.052304, BabakPRL2020,BoundStateLuca}. Striking collective behavior emerges in large collections of chemically active colloids~\cite{Jaime2019,ouazan2021non,OuazanReboul2023,VincentPRL} involving chasing dynamics. Continuum theories that minimally capture the essence of nonreciprocal interactions in scalar mixtures have been proposed~\cite{NRCH,you2020nonreciprocity} and are being explored intensely~\cite{ThieleFirst,ThieleHuelsmann,saha2022effervescent}. In polar active matter, with orientation as a relevant degree of freedom, non-reciprocity can be incorporated in a multitude of ways - directly in the alignment rules~\cite{fruchart2021non,Stark_Knežević2022}, through a dependence of the spin-spin interactions on spatial anisotropy~\cite{LokrshiMaitraPRE,SAM_VisionCone,NRswings_PRL}, or through quorum sensing~\cite{BenoitYu2023}. Explorations of the collective behavior of systems with polar, nematic, or chiral order constitute an active sub-field of research \cite{fruchart2021non, Klapp_Loos_2020, Sinha_2024}. Novel steady states arise in all the examples mentioned here due to the simultaneous breaking of parity and time reversal symmetry leading to chiral motion in polar mixtures~\cite{fruchart2021non,Uchida2010_PRL}, traveling waves in scalar mixtures~\cite{NRCH}, and stress-strain cycles in odd solids~\cite{Scheibner2020,ActiveSolidBartolo} relating this class of phenomenon to odd response~\cite{Shankar2022}.

A few examples of experimental systems that exhibit nonreciprocal interactions are - active Janus colloids~\cite{Ramin-Liverpool-Ajdari2005, RaminHowsePhysRevLett.99.048102,SmallSharan2023}, light actuated colloids with a vision cone~\cite{Lavergne2019BechingerGroup,Durve2018VisionConeSimulations}, and dusty plasma~\cite{DustyPlasma}. In this work, we will focus on Janus colloids, where nonreciprocal interactions between densities and polarities are realized through chemical field-mediated interactions. The nonequilibrium active dynamics of self-propelled Janus colloids are due to self-phoresis~\cite{Ramin-Liverpool-Ajdari2005}, which harnesses the force-free mechanism of diffusiophoresis at microscopic lengthscales~\cite{Anderson1989,Golestanian2019phoretic}. These particles are able to catalyze a chemical reaction on their surface which modifies the density profile of the involved reactants and products~\cite{Kapral_PhysRevLett.98.150603}. Depending on the geometrical properties of the colloid, changes in these chemical substrates' concentration may lead to the particle net motion~\cite{golestanian2007designing}. Moreover, the coupling of different colloids to the same substrates induces an effective long-ranged interaction among different particles.

The versatility of a system of Janus colloids arises from the variety in their dynamical response~\cite{PhysRevE.89.062316,Ramin-LesHouches2018}. In a collection of identical colloids, possible collective dynamics include system-wide phase separation, pattern formation with a selected lengthscale, and oscillations similar to the Jeans instability in gravitational systems as explored in \cite{PhysRevE.89.062316}. Numerical solutions of the dynamics for chemorepulsive colloids in \cite{Liebchen2015, liebchen2017phoretic} show that the linear instabilities indeed pave the route to dynamic aggregates, and spatiotemporal patterns including traveling waves. For two interacting Janus colloids, the interplay of orientational dynamics, self-propulsion, and drift produces a complex effective potential landscape leading to bound orbits with internal jiggling and chiral bound state~\cite{saha2019pairing}.

In this paper, we have studied the collective behavior of two species of phoretic colloids starting from their microscopic dynamics and building the continuum field theories. In section \ref{sec:model}, we introduce the model of multi-species Janus colloids. In section \ref{sec:twospecies}, we discuss the linear stability analysis of the two-species case, while in appendix \ref{app:OneSpecies} the one-species one. Finally in section \ref{sec:twospeciesrelaxed}, we study what happens to the two-species system whenever the polarization fields relax fast enough so that they can be adiabatically eliminated. 

\begin{figure}
\centering
\includegraphics[width = 0.95\linewidth]{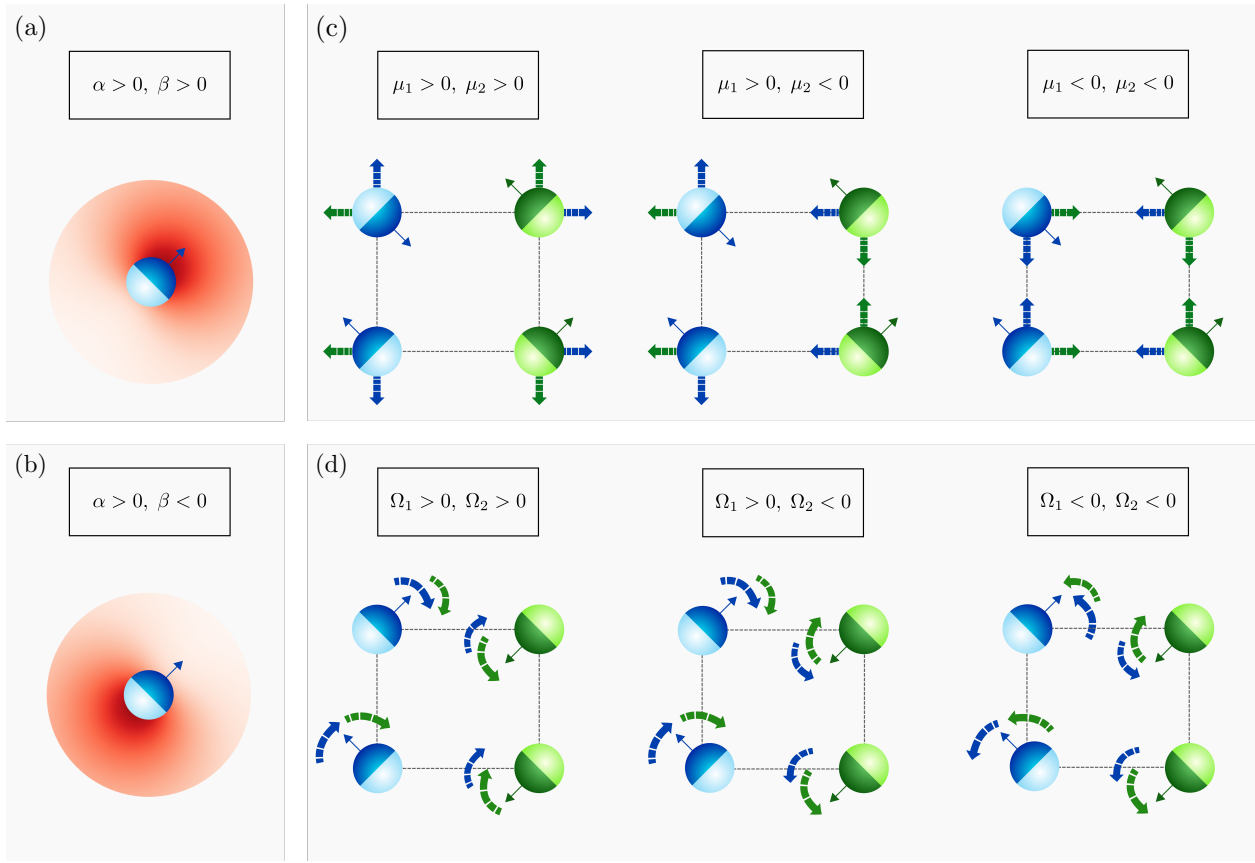}
\caption{Schematic showcasing the complexity in the dynamics that can be harnessed to tune the system to the collective behavior of choice. (a) shows a Janus colloid that produces a chemical field (profile shown by red heatmap) with rate $\alpha>0$. $\beta$ determines the asymmetry in production along the polarity $\bm{n}$. (b) is similar to (a) but with the sign of $\beta$ reversed thus flipping the profile of the chemical field in (a). (c) lists all possible types of center-to-center interaction between two different species of Janus colloids (blue and green) due to chemotactic drift. To fix ideas, we illustrate pairwise interactions only, and the two involved particles are joined by a dotted black line. For both $\mu_{a}$ positive (negative), all interactions are repulsive (attractive). For $\sign(\mu_1) \neq \sign({\mu}_2)$, nonreciprocal interactions emerge where one species chases the other. (d) shows the effect of chemotactic alignment: whenever $\Omega_a$ is positive (negative), particles re-orient towards (away from) high-density regions resulting in a novel form of orientational order. Orientational dynamics leads to effective pairwise attraction-repulsion or chasing depending on the sign of $\Omega_a$.  
}\label{fig:schematic}
\end{figure}

\section{Multi-species Janus particles}\label{sec:model}

We study the dynamics of $n$ different species of Janus colloids \cite{kanso2019phoretic,PhysRevE.89.062316} interacting with a chemical substrate. These particles are sensitive to the spatial gradient of the substrate and respond accordingly in two different ways: (i) their velocity varies proportionally to the gradient of the chemical concentration (chemotactic drift), (ii) they re-orient along it (chemotactic alignment). Moreover, each Janus particle contributes to the production (or consumption) of the substrate particles. This inhomogeneity in the chemical in the proximity of the Janus particle's surface produces a slip velocity, which induces self-propulsion in the direction of the particle's axis \cite{golestanian2007designing}.
An effective nonreciprocal interaction between the two species follows from the direct interplay with the chemical.
We describe the system by looking at the 3-dimensional dynamics of Janus colloids of species $a\in\{1,2,\dots, n\}$ with position $\bm{r}_a$ and orientation $\bm{n}_a$ which specifies the direction of self-propulsion. They evolve in time according to the Langevin equations
\begin{equation}\label{eq:langevin}
\begin{aligned}
\frac{\mathrm{d}\bm{r}_a}{\mathrm{d}t}&=-\mu_a\nabla c+v_a\bm{n}_a+\bm{\xi}_a,\\
\frac{\mathrm{d}\bm{n}_a}{\mathrm{d}t}&=\bm\omega_a\times \bm{n}_a\\
&=(\Omega_a \bm{n}_a\times\nabla c+\bm{\zeta}_a)\times \bm{n}_a\\
&=\Omega_a\left(\mathbb{I}-\bm{n}_a\bm{n}_a\right)\cdot\nabla c+\bm{\zeta}_a\times\bm{n}_a,
\end{aligned}
\end{equation}
here $\mathbb{I}$ denotes the 3-dimensional identity matrix, and $v_a$ the self-propulsion speed. The sign of the chemotactic mobility $\mu_a$ prescribes whether the particle moves following increasing ($\mu_a<0$) or decreasing ($\mu_a>0$) gradients of the substrate density field $c(\bm{r},t)$. Similarly, the coefficient $\Omega_a$ leads to alignment ($\Omega_a>0$) or anti-alignment ($\Omega_a<0$) of the particle velocity along the gradient $\nabla c(\bm{r},t)$. In writing equation~\eqref{eq:langevin} we ignore terms that are quadratic in $\bm{n}_a$, keeping only those that are leading order in $\bm{n}_a$. This simplification is consistent with ignoring nematic order during coarse-graining as we will discuss later; we refer to appendix  \ref{app:CG} for more details. The fluctuating nature of equation \eqref{eq:langevin} is encoded in the Gaussian white noises $\bm{\xi}_a(t)$ and $\bm{\zeta}_a(t)$, characterized by zero mean and variance
\begin{equation}
\langle \zeta_{a,i}(t)\zeta_{b,j}(t)\rangle=2D_{r,a}\delta_{ab}\,\delta_{ij}\,\delta(t-t'),\;\;\text{and}\;\;\langle \xi_{a,i}(t)\xi_{b,j}(t)\rangle=2D_a\delta_{ab}\,\delta_{ij}\,\delta(t-t'),
\end{equation}
where the labels $i,\,j$ identify the 3-dimensional space components of the vectors. 
Note that the conservation of the modulus $|\bm{n}_a|^2=1$ from equation \eqref{eq:langevin} implicitly assumes the Stratonovich representation of the stochastic differential equation \cite{gardiner2009stochastic,pavliotis2014stochastic}.
The set of dynamical equations~\eqref{eq:langevin} is completed by including the evolution of the substrate density field  $c(t)$, which is given by
\begin{equation}\label{eq:c}
\partial_t c-D_c(\nabla^2 -\kappa^2)c=\sum
\left[\alpha_a\rho_a-\beta_a\nabla\cdot\bm{p}_a\right].
\end{equation}
The fields $\rho_a(\bm{r},t)$ and $\bm{p}_a(\bm{r},t)$ denote respectively the particle density and the polarization field associated with the species $a$, and they are defined as follows
\begin{equation}\label{eq:RhoPolar}
\rho_a = \Big\langle\sum_i \delta(\bm{r} - \bm{r}_{a,i})\Big\rangle,\;\;\;\; \bm{p}_a = \Big\langle\sum_i \bm{n}_{a,i} \delta(\bm{r} - \bm{r}_{a,i})\Big\rangle,
\end{equation}
where $i$ runs over all particles of the $a-$th species, and the average is evaluated with respect to the noise realizations. As clear from equation \eqref{eq:c}, the substrate diffuses with diffusion constant $D_c$ and degrades exponentially over the timescale $1/D_c \kappa^2$, associated with the screening lengthscale $\kappa^{-1}$. We also assume that Janus particles constitute point-like sinks or sources for the substrate. If the sign of the parameter $\alpha_a$ in equation \eqref{eq:c} is positive (negative) the particles of species $a$ produce (consume) the substrate. Similarly, the dipole term $\beta_a$ accounts for the head-tail asymmetry in the production (or consumption) along $\bm{n}_a$. If $\beta_a>0$ the Janus particles produce more (or consume less) substrate around the catalytic cap of $\bm{n}_a$, while they produce less (or consume more) otherwise. We represent this in figure \ref{fig:schematic} (a)-(b), where, the net production of the chemical ($\alpha>0$) is modulated along its symmetry axis by the sign of $\beta$. In figure \ref{fig:schematic} (c), we consider the joint effect of the production of the chemical and chemotaxis for two different species of Janus colloids, marked by the two different blue and green colors. In particular, we assume the scenario represented in figure \ref{fig:schematic} (a), where the chemical gradient grows in the self-propelling direction. If we consider the interactions to be pairwise, given a Janus particle in the picture, the bold blue and green arrows identify the net qualitative force due to the colloid of the same color along the dotted line. In the case where $\mu_{1,2}>0$, the particles, being a source of chemical, are effectively repelled by each other. If $\mu_{2}<0$ changes sign, green colloids are attracted by other particles, while blue continue to be repelled: this is a prototypical manifestation of effective nonreciprocal interactions. In the last case, being $\mu_{1,2}<0$, the particles are all effectively attracted by each other. Similarly, in figure \ref{fig:schematic} (d), we represent the effect of pairwise chemotactic alignment for the two species of particles. Whenever $\Omega_{1,2}>0$ the particles tend to align following higher gradients of the chemical concentration $\nabla c$, which in this example coincides with the self-propelling direction (tiny arrow), while they anti-align along it otherwise.

Our goal is to characterize the various dynamical steady states by looking at the effect of chemically mediated interactions on the particle distribution $\rho_a$ and orientation $\bm{p}_a$. It follows from equation~\eqref{eq:RhoPolar} that $\bm{p}_a$ and $\rho_a$ are coupled fields, for instance, $\bm{p}_a$ vanishes at a point where $\rho_a$ is zero.
As we discuss in \ref{app:CG}, $\rho_a$ and $\bm{p}_a$ are respectively the zeroth and first moment in $\bm{n}_a$ of the joint probability density $\mathcal{P}_a = \langle\sum_i \delta(\bm{r} - \bm{n}_{a,i})\delta(\bm{r} - \bm{r}_{a,i})\rangle$. In general, one can construct an infinite hierarchy of equations where the time evolution of the $m-$th moment of $\bm{n}$ depends on the $m'-$th moment,  with $m'\ge m$. In the spirit of describing the collective behavior and keeping only the most relevant fields, i.e., those reflecting conservation laws or broken symmetry in the system, we restrict our analysis to the first and second moments of the orientation $\bm{n}$ by truncating the corresponding hierarchy of infinitely many equations in the $\bm{n}$ moment expansion~\cite{ThermalActiveColloids}; the details of the derivation are reported in appendix \ref{app:CG}. Thus, we get equations for the coarse-grained fields $\rho_a$ and $\bm{p}_a$ 
\begin{equation}\label{eq:rho}
\partial_t\rho_a=-\nabla\cdot\left(v_a\bm{p}_a-\mu_a\rho_a\nabla c\right)+D_a\nabla^2\rho_a ,
\end{equation}
\begin{equation}\label{eq:ptruncated}
\partial_t\bm{p}_a=(-2D_{r,a} + D_a\nabla^2)\bm{p}_a+\mu_a\nabla\cdot\left(\bm{p}_a\nabla c\right) +\frac{2}{3}\Omega_a\rho_a\nabla c -\frac{v_a}{3}\nabla\rho_a,
\end{equation}
that allow us, together with equation \eqref{eq:c}, to describe the macroscopic behavior of the system under the assumption of negligible nematic order parameter $\mathbb{Q}_a = \langle\sum_i (\bm{n}_{a,i}\bm{n}_{a,i}-\mathbb{I}/3)\delta(\bm{r} - \bm{r}_{a,i})\rangle$. Equation \eqref{eq:rho} describes the conserved evolution of the particle density $\rho_a$ via three different contributions: the first accounts for advection of particles because of self-propulsion, the second is the effect of chemotaxis, and the third is translational diffusion.
Similarly, in equation \eqref{eq:ptruncated} for the polar field $\bm{p}_a$, the first term represents the effect of orientational and translational diffusion. By its very definition in equation~\eqref{eq:RhoPolar}, $\bm{p}_a$ is coupled with $\rho_a$ -- consequently, both diffusion and phoretic drift of $\rho_a$ affect $\bm{p}_a$ through the second and the third terms respectively. To the lowest order in spatial gradients, $\bm{p}_a$ rotates to align with the local substrate gradient, an effect that is encoded in the third term proportional to the coefficient of alignment $\Omega_a$.  Finally, the last term can be interpreted as a pressure term that measures how self-propulsion influences local order.

\subsection{Linearized dynamics}\label{sec:linearizeddynamics}
In general, equations \eqref{eq:rho}, \eqref{eq:ptruncated}, complemented with equation \eqref{eq:c} constitute a set of $2n+1$ nonlinear partial differential coupled equations for an equal number of scalar and vector fields. A great reduction of their complexity is achieved by studying their linearized form, by looking at perturbations with respect to a simple and physically relevant solution. We perform linear stability analysis of equations~\eqref{eq:c}, \eqref{eq:rho}, and \eqref{eq:ptruncated} by considering small perturbations around the spatially homogeneous solution
\begin{equation}\label{eq:homsol}
   \rho_a=\bar{\rho}_a,\quad\bm{p}_a=0,\quad c=\bar{c}=\frac{\sum_a\alpha_a\bar{\rho}_a}{D_c\kappa^2},
\end{equation}
corresponding to a state where the $n$ species are well mixed, i.e., there are modulations in the density, and no orientational order is present. The stationary substrate density $\bar{c}>0$ is constant, as its net production by Janus colloids is balanced by its degradation rate. We are interested in solutions that are perturbations of equation \eqref{eq:homsol} in the form of
\begin{equation}\label{eq:linsol}
\begin{aligned}
\rho_a(\bm{r},t)=\bar{\rho}_a+\delta\rho_a(\bm{r},t),\quad
\bm{p}_a(\bm{r},t)=\delta\bm{p}_a(\bm{r},t),\quad
c(\bm{r},t)=\bar{c}+\delta c(\bm{r},t).
\end{aligned}
\end{equation}
We can now build the dynamical equations for $\delta \rho_a$, $\delta\bm{p}_a$ and $\delta c$ by discarding contributions that are  higher than the linear order in the perturbations. This procedure leads to the following set of linearized equations
\begin{equation}\label{eq:rhoplinear}
\begin{aligned}
&\partial_t\delta\rho_a(\bm{r},t)=\mu_a\bar{\rho}_a\nabla^2\delta c(\bm{r},t)+D_a\nabla^2\delta\rho_a(\bm{r},t)-v_a\nabla\cdot\delta\bm{p}_a(\bm{r},t),\\
&\partial_t\delta \bm{p}_a(\bm{r},t)=-\frac{v_a}{3}\nabla\delta\rho_a(\bm{r},t)+\frac{2}{3}\Omega_a\bar{\rho}_a\nabla \delta c(\bm{r},t)+(D_a\nabla^2-2D_{r,a})\delta\bm{p}_a(\bm{r},t),\\
&\partial_t\delta c(\bm{r},t)=-D_c\left(\kappa^2- \nabla^2\right) \delta c(\bm{r},t)+\sum_{a}\left[\alpha_a\delta\rho_a(\bm{r},t)-\beta_a\nabla\cdot\delta\bm{p}_a(\bm{r},t)\right].
\end{aligned}
\end{equation}
The system of equations~\eqref{eq:rhoplinear} represents the starting point to discuss the linear stability of the disordered state and the onset of order in the system. Although linear, the current form of equations \eqref{eq:rhoplinear} is still very complex due to the large number of fields involved and the related parameters. 
A further simplification follows from the physical assumption that the deviations of the substrate density $\delta c$ from the space homogeneous solution relax much faster than those of the density and polarization field of the Janus particles, that is $\partial_t \delta c(\bm{r},t)\simeq0$, or equivalently
\begin{equation}\label{eq:deltacstat}
    D_c\left(\kappa^2- \nabla^2\right) \delta c(\bm{r},t)=\sum_{a}\left[\alpha_a\delta\rho_a(\bm{r},t)-\beta_a\nabla\cdot\delta\bm{p}_a(\bm{r},t)\right].
\end{equation}
Furthermore, it is convenient to express  the time evolution of $\drak$, $\dpak$, and $\dck$ in terms of their Fourier modes, whose dynamics directly follow from equations \eqref{eq:rhoplinear} and \eqref{eq:deltacstat} as  
\begin{equation}\label{eq:FTeq}
\begin{aligned}
&\partial_t\drak=-k^2\left[\mu_a\bar{\rho}_a\dck+D_a\drak\right]-v_a (i\bm{k})\cdot\dpak,\\
&\partial_t \dpak=-\frac{i\bm{k}}{3}\left[v_a\drak-2\Omega_a\bar{\rho}_a \dck\right]-(D_ak^2+2D_{r,a})\dpak,\\
&D_c\left(\kappa^2+k^2\right)\dck= \sum_{a}\left[\alpha_a\drak-\beta_a(i\bm{k})\cdot\dpak\right];
\end{aligned}
\end{equation}
as a convention the Fourier transform of a given function $f(\bm{r})$ reads  $f(\bm{k})\equiv \int_{-\infty}^{+\infty}\mathrm{d}\bm{r}\,e^{-i\bm{k}\cdot\bm{r}}\,f(\bm{r})$. For the sake of completeness, we mention that the closure of the set linearized equations at the nematic field $\mathbb{Q}_a$ order would simply lead to renormalization of $D_{r,a}$ to $D_{r,a}+k^2v_a^2/(45D_{r,a})$, thus leaving the qualitative behavior of the system unaltered.

We can reduce the degrees of freedom of the problem by looking at the transverse and longitudinal components of $\dpak$. The initial value of the perturbation $\dpak$ in the direction transverse to the wave vector $\bm{k}$ decays with relaxation time $D_{r,a} + D_a k^2 $, which remains positive and finite at all wave numbers. Moreover, the dynamics of $\drak$ and $\dck$ in equations \eqref{eq:FTeq} depends on those of $\dpak$ only via $i\bm{k}\cdot \dpak$, i.e., the divergence of the polar field in real-space coordinates. Accordingly, we decompose the polar field along $\hat{\bm{k}}=\bm{k}/k$ and in the transverse direction as
\begin{equation}\label{eq:decomposition}
\begin{aligned}
    \dpak&=(\hat{\bm{k}}\cdot\dpak)\,\hat{\bm{k}}+(\mathbb{I}-\hat{\bm{k}}\hat{\bm{k}})\cdot\dpak\\
    &=\dwak\,\hat{\bm{k}}+\dperak,
\end{aligned}
\end{equation}
where $\dwak$ is the parallel (longitudinal) component of $\dpak$ to $\hat{\bm{k}}$ and $\dperak$ is a vector which belongs to the plane perpendicular to $\hat{\bm{k}}$ (transverse component). As anticipated, one can check that the perpendicular component $\dperak$ of the polar field is decoupled from the other fields and it relaxes exponentially in time according to $\partial_t \dperak=-(D_ak^2+2D_{r,a})\dperak$.
Finally, by substituting the explicit expression of $\dck$ in equations \eqref{eq:FTeq} for $\drak$ and $\dpak$, the dynamics of the system are encoded in those of the two slow (scalar) fields

\begin{equation}\label{eq:FTlineqtime}
\begin{aligned}
    &\partial_t
    \drak
  =\sum_b\left[k^2\mathcal{G}_{ab}^{tt}(k)\,\drbk+ik\,\mathcal{G}^{tr}_{ab}(k) \,\dwbk\right],\\[3pt]
  &\partial_t \dwak=\sum_b\left[ik\, \mathcal{G}^{rt}_{ab}(k)\,\drbk+\mathcal{G}_{ab}^{rr}(k)\,\dwbk\right],
\end{aligned}
\end{equation}
where we have defined the $k-$dependent matrices of the inter-species couplings
\begin{small}
\begin{equation}\label{eq:dynamicalmatrices}
\begin{split}
    &\mathcal{G}_{ab}^{tt}(k)\equiv -D_a\delta_{ab}-\mu_a\bar{\rho}_a\alpha_b/[D_c(\kappa^2+k^2)],\\
    &\mathcal{G}^{rt}_{ab}(k)\equiv  -\delta_{ab}v_a/3+ 2\Omega_a\bar{\rho}_a\alpha_b/[3D_c(\kappa^2+k^2)],
\end{split}\quad\quad
\begin{split}
  &\mathcal{G}^{tr}_{ab}(k)\equiv  -v_a\delta_{ab}+\mu_a\bar{\rho}_a\beta_b \,k^2/[D_c(\kappa^2+k^2)],\\  
  &\mathcal{G}_{ab}^{rr}(k)\equiv -(D_ak^2+2D_{r,a})\delta_{ab}+2\Omega_a\bar{\rho}_a\beta_b\, k^2/[3D_c(\kappa^2+k^2)].
\end{split}
\end{equation}
\end{small}
All the elements of these interaction matrices have a similar structure 
\begin{equation}\label{eq:structure}
\mathcal{G}_{ab} = \mbox{single particle dynamics}+ \mbox{mobility coefficient for species }a \times \mbox{activity coefficient for species }b.
\end{equation}
The dynamical matrices in equations \eqref{eq:dynamicalmatrices} are not symmetric, which means that the effective interactions between the various species, following from the different coupling with the chemical substrate, are nonreciprocal. Note that while the homogeneous production/consumption rate $\alpha_a$ contributes at all scales, the dipole contribution proportional to $\beta_a$ is expected to be sub-leading in the macroscopic limit $k\rightarrow 0$, as it scales as $k^2$.  

In the next two sections, we present a stability analysis of the two-species system in two particularly significant cases. In the first we consider the fully coupled dynamics focusing mainly on the simpler case where the bare translational and rotational diffusivities of the species are identical. In this case, the dynamics reduce to the coupled dynamics of a conserved and a non-conserved field which can undergo an instability to a spontaneously oscillating state. In the second case, we assume that the timescales $D_{r,a}^{-1}$ are smaller than all other timescales in the system, such that the polarization fields can be enslaved to the density ones to obtain the equations for two active densities, representing a microscopic realization of the Nonreciprocal Cahn-Hilliard model~\cite{NRCH}.

\section{Coupled dynamics of two species}\label{sec:twospecies}

Here we simplify our analysis by restricting to the case of two species of Janus colloids. This represents the simplest, yet physically relevant, system of Janus particles that display nonreciprocal interactions due to phoretic coupling between similar types of fields - the density fields or the polar fields. To reduce the large parameter space, we first assume equal diffusivity $D_1=D_2=D$, orientational noise timescale $D_{r,1}=D_{r,2}=D_r$ for the two species, and equal self-propelling velocity $v_1=v_2=v$, while allowing the phoretic mobility coefficients to be different. The dynamics of the system are described by the spectral properties of the full dynamical matrix $\mathcal{G}$, which is here defined via equation \eqref{eq:dynamicalmatrices} as 
\begin{equation}\label{eq:D2s}
    \mathcal{G}(k)=\left(
\begin{array}{cc}
k^2\,\mathcal{G}_{ab}^{tt}(k) & ik\,\mathcal{G}_{ab}^{tr}(k) \\[5pt]
ik\,\mathcal{G}_{ab}^{rt}(k) & \mathcal{G}_{ab}^{rr}(k)\\
\end{array}\right).
\end{equation}
In particular, the eigenvalues of $\mathcal{G}$ quantify the rate with which the different modes in the linearized system of equations grow or decay exponentially with time. In addition, complex eigenvalues imply an oscillatory response to perturbations. Here we list the four eigenvalues $\Lambda_i(k)$ with $i=1-4$ of the $\mathcal{G}$ matrix. The first pair of eigenvalues $\Lambda_{1,2}$ are
\begin{equation}\label{eq:eigs1}
    \Lambda_{1,2}(k) = -\left(D_r+k^2D\pm\sqrt{\Delta_1(k)}\right),\quad\quad 
    \Delta_1(k)=\left(D_r\right)^2-\frac{k^2v^2}{3},
\end{equation}
and they are associated with stable modes that are independent of the phoretic effects.
The second pair of eigenvalues $\Lambda_{3,4}$ are
\begin{equation}\label{eq:eigs2}
    \begin{aligned}
& \Lambda_{3,4}(k) = -\left(D_r+Dk^2+\frac{K(k) \bar{\rho}}{2}\frac{\bm{\alpha}\cdot\bm{\mu}-\bm{\beta}\cdot\bm{\Omega}}{D_c}\pm \sqrt{\Delta_2(k)}\right), \\
&   \Delta_2(k)
    = \left( \bar{\rho} K(k)\frac{\bm{\alpha}\cdot\bm{\mu}-\bm{\beta}\cdot\bm{\Omega}}{2D_c}\right)^2+\frac{ \bar{\rho} K(k)}{D_c}\left[v\left(\bm{\alpha}\cdot\bm{\Omega}+\frac{k^2}{3}\bm{\beta}\cdot\bm{\mu}\right)-D_r\left(\bm{\beta}\cdot\bm{\Omega}+\bm{\alpha}\cdot\bm{\mu}\right)\right]+\Delta_1(k),
    \end{aligned}
\end{equation}
where we define the screening parameter $K(k)=k^2/(k^2+\kappa^2)$, total average density $\bar{\rho}=\bar{\rho}_1+\bar{\rho}_2$, and we group the phoretic coefficients for the two species as pairs of numbers which we write in a compact manner adapting the vector notation $\bm{\alpha}=(\alpha_1, \alpha_2)$, $\bm{\beta}=(\beta_1,\beta_2)$, $\bm{\mu}= (\bar{\rho}_1\mu_1 \bar{\rho}^{-1}, \bar{\rho}_2\mu_2 \bar{\rho}^{-1})$, $\bm{\Omega}=(\bar{\rho}_1\Omega_1 \bar{\rho}^{-1}, \bar{\rho}_2\Omega_2 \bar{\rho}^{-1})/3$. The factor $K(k)$ is determined by the relative magnitude of $k/\kappa$ - at fixed $\kappa$ it vanishes when $k \ll \kappa$ and approaches unity when $k\gg\kappa$. 
We recall that the linear regime relies on the condition of steady chemical density $\bar{c}=(\bar{\rho}_1\alpha_1+\bar{\rho}_2\alpha_2)/(D_c \kappa^2)\ge 0$, which is obtained by the balance of net positive production of the substrate via the activity of the Janus colloids and its spontaneous degradation, expressed by $\kappa>0$. 
Contrarily to the eigenvalues $\Lambda_{1,2}$, the pair $\Lambda_{3,4}$ depends on the phoretic coefficients. This implies that the mode structure of the linearized dynamics splits into two parts. The first set of modes $\Lambda_{1,2}$ describes the linear response of a self-propelled polar field coupled to a density field and no other source of activity or interaction. On the other hand, $\Lambda_{3,4}$ receives contributions from groups of phoretic parameters whose relative magnitudes can be tuned to give rise to linear instabilities. The reason behind this partition at the linear level is the consideration of a single chemical species $c$ which leads to the factorized structure in the dynamical matrix discussed in equations~\eqref{eq:dynamicalmatrices} and \eqref{eq:structure}. We will show later in this section that this decomposition does not hold if the diffusivities and the self-propelling velocities of the two species are unequal. For $k \ll \kappa$, as $K(k)$ approaches the ratio $k^2/\kappa^2$, and the pair $\Lambda_{3,4}$ approaches $\Lambda_{1,2}$, i.e. phoretic effects are suppressed in dynamics occurring at lengthscales much larger than $\kappa^{-1}$. 

\begin{figure}
\centering
\includegraphics[width = \linewidth]{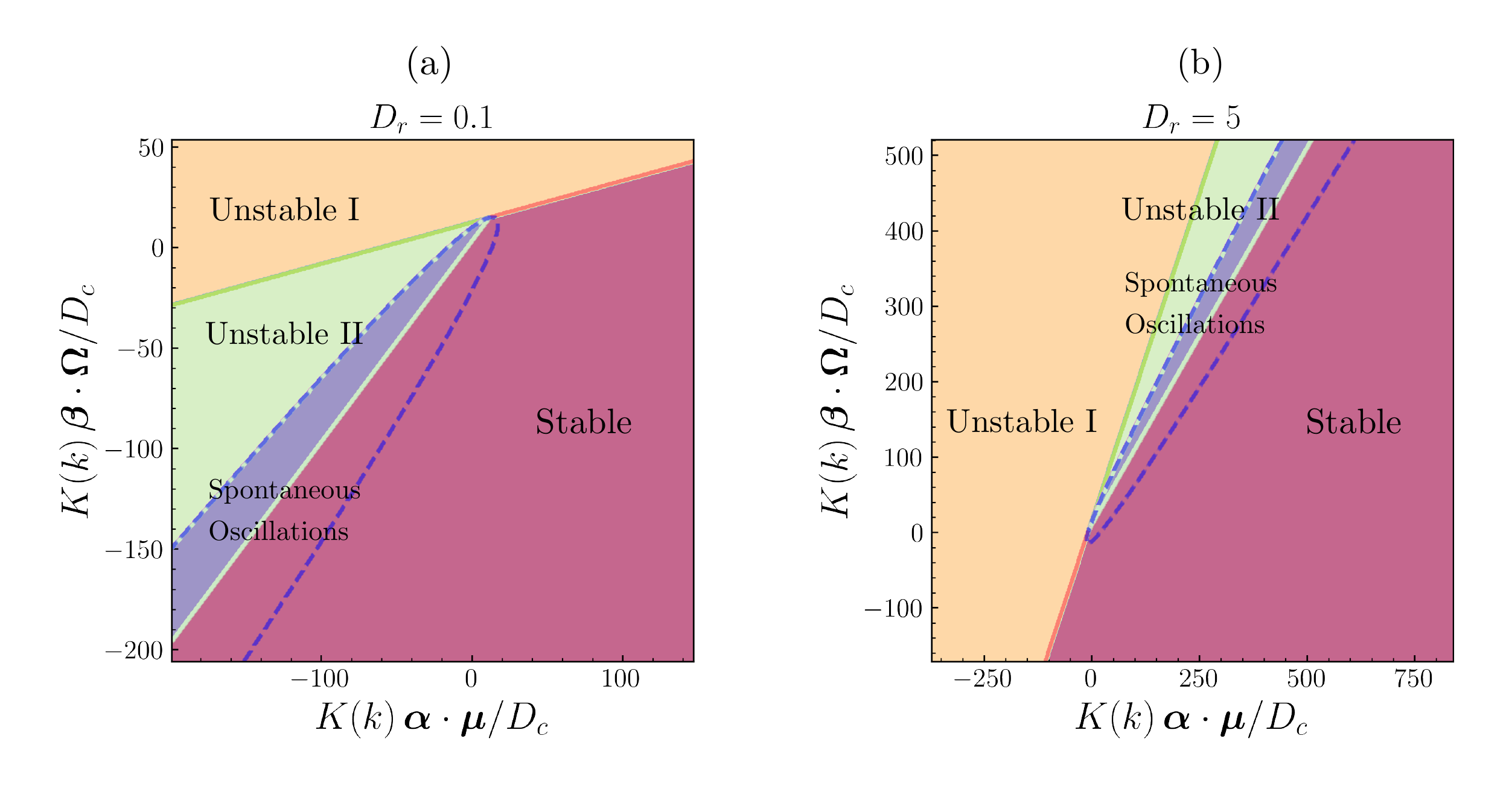}
\caption{
Phase diagrams for $D_1=D_2=D=1$, $v_1=v_2=v=2.5$ for two different values of $D_{r,1}=D_{r,2}=D_r$, and $k/\kappa =1.5$. The magenta area is associated with a homogeneous phase. The dashed blue line identifies exceptional points: it separates real and complex oscillating modes, that, if unstable, could lead to ``Spontaneous Oscillations" (blue area). In the ``Unstable I" region (orange area) the conserved mode is unstable and the non-conserved mode is stable; it becomes unstable in the ``Unstable II" phase. 
(a) corresponds to small values of  $D_r$. Assuming  $\alpha_a>0$ and  $\beta_a>0$, the homogeneous phase is stable for $\bm\alpha\cdot\bm\mu>0$, i.e., for effective repulsive interaction among particles. For $\bm\alpha\cdot\bm\mu<0$ and $\bm\beta\cdot\bm\Omega<0$, the colloids experience attractive interaction and re-orient away from high-density regions: this could lead to a homogeneous phase whenever the latter dominates the former, to spontaneous oscillations if they are counterbalanced, or to the creation of asters otherwise. For $\bm\beta\cdot\bm\Omega>0$ and $\bm\alpha\cdot\bm\mu<0$, both chemotactic drift and alignment tend to aggregate particles, thus leading to an instability of the density mode. In (b) $D_r$ takes larger values. 
The main difference from (a)  consists in the fact that a locally ordered phase can arise only for strong enough aligning interaction $\bm{\beta}\cdot\bm{\Omega}>0$ compared with $D_r$.}
\label{fig:PDtwospecieslikeone}
\end{figure}

We will now discuss the instabilities indicated by $\Lambda_i$ and how they can be tuned. We keep in mind that the behavior at small $k$ has the most predictive power in determining the nonequilibrium steady states exhibited by the system. The eigenvectors corresponding to $\Lambda_{i}$ contain information about the combination of fields whose perturbations show exponential growth. However, the complexity of the problem allows us to make only the following comment. At vanishing $k$, the system is described by two conserved modes ($\Lambda_{1,3}$) (a combination of the two density fields) and two non-conserved modes ($\Lambda_{2,4}$) (combination of polarization fields). This point will be clear from the Taylor expansion in $k$. An exponential growth in the conserved modes could lead to active phase separation. The growth of fluctuations in the non-conserved modes, presumably leading to a growth in the longitudinal part of $\bm{p}$ leads to a relatively unexplored type of orientational order called asters~\cite{LokrshiMaitraPRE}. 

We start by showing that $\Lambda_{1,2}$ is always stable at small $k$. Taylor expanding $\Lambda_{1,2}$ we find
\begin{equation}\label{eq:smallq13}
\begin{aligned}
    &\Lambda_1(k)= -2D_r-k^2\left(D-\frac{v^2}{6D_r}\right)+O(k^4),\\ &\Lambda_2(k)=-k^2 D^{\rm eff}(0)+O(k^4),
\end{aligned}
\end{equation}
where we introduce the effective diffusivity modified by dynamics of the polarization field as for an active Brownian particle, $D^{\rm eff}(k)\equiv D+v^2/[3(2D_r+Dk^2)]$ \cite{romanczuk2012active, stenhammar2014phase, siebert2018critical, solon2015active}. $D^{\rm eff}(k)$ is always greater than zero meaning that density perturbations eventually decay. Note however that the polarization field can show patterning at values of $k \approx 2D_r (D - v^2/6D_R)^{-1}$.

\begin{figure}
\centering
\includegraphics[width = 0.75\linewidth]{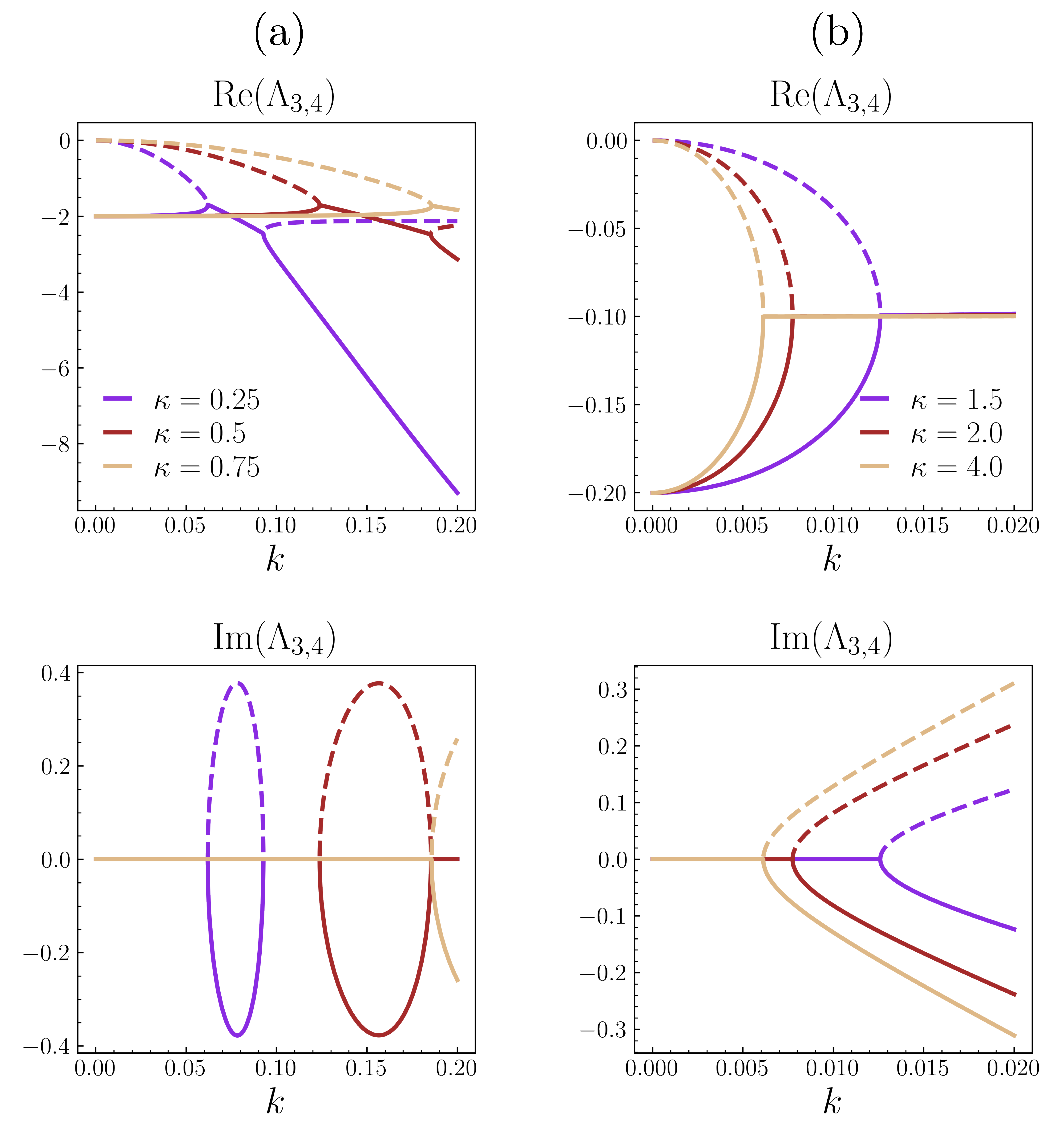}
\caption{Here we present two instances of the $\kappa$-dependence of the eigenvalues $\Lambda_{3,4}$ in equation \eqref{eq:eigs2} for the two-species case. In (a), corresponding to the small-$v$ limit, as $\kappa$ decreases, the system displays oscillations at decreasing values of $k$. The parameters are here chosen to be $D = D_r = 1$
$D_c = 0.5$
$v = 0.1$
$\rho_1=5$, $\rho_2 =1$,
$\alpha_1=1.5$, $\alpha_2 = 2.5$, 
$\beta_1=0.5$, $\beta_2 = 0.25$, 
$\mu_1=1$, $\mu_2 = 2$,
$\Omega_1=0.5$, $\Omega_2 = 1.4$. In (b), corresponding to the large $v$-limit, as the effect of screening increases oscillations occur at larger values of $k$. The associated  parameters are $D = D_r = 0.1$
$D_c = 1$
$v = 30$
$\rho_1=5$, $\rho_2 =1$,
$\alpha_1=1.5$, $\alpha_2 = 2.5$, 
$\beta_1=2.5$, $\beta_2 = 5.25$, 
$\mu_1=1$, $\mu_2 = 2$,
$\Omega_1=5.5$, $\Omega_2 = 5$. }\label{fig:Eigtwo}
\end{figure}

\subsection{Instabilities in $\Lambda_{3,4}$}\label{sec:lambda34}

The eigenvalues $\Lambda_{3,4}$ depend on all possible pairings of phoretic mobilities and chemical activity which are four in total. $\bm{\alpha} \cdot \bm{\mu}$ and $\bm{\beta} \cdot \bm{\mu}$ arise from particle drift in response to isotropic and anisotropic chemical production respectively. Similarly, $\bm{\alpha} \cdot \bm{\Omega}$ and $\bm{\beta} \cdot \bm{\Omega}$ quantify alignment with substrate gradient whether isotropic and anisotropic respectively.
We first show a stability diagram spanned by $K(k) \bm{\alpha} \cdot \bm{\mu}/D_c$ and $K(k) \bm{\beta} \cdot \bm{\Omega}/D_c$; the details about the construction of the phase diagram are discussed in appendix \ref{app:conpd}. The phase diagram assumes a fixed value of $k$, however, its structure is qualitatively the same for a generic choice of the wave vector. Both $\Lambda_{3,4}$ are positive in the ``Unstable II" green area in figure~\ref{fig:PDtwospecieslikeone}, corresponding to an instability in both the density and the polar fields, presumably a state with phase separation coupled with aster formation. The ``Unstable I" orange area refers to the instability of the conserved mode only, resulting in phase separation driven by phoretic activity. These two unstable regions are separated by the line along which $\Lambda_3=0$, and $\Lambda_4<0$. Here, $\Lambda_{3,4}$ are complex numbers for $\Delta_2<0$, which can happen only when $(\bm\alpha\cdot\bm\mu-\bm\beta\cdot\bm\Omega)<0$. The homogeneous phase is ``Stable" in the magenta region of figure~\ref{fig:PDtwospecieslikeone} case of incoherent self-propulsion ($\bm\beta\cdot\bm\Omega<0$) and chemotactic repulsion ($\bm\alpha\cdot\bm\mu>0$). The boundary (dashed blue line) between the green and the blue regions is made of exceptional points:  they separate the unstable complex, associated with ``Spontaneous Oscillations", and unstable real modes. The blue and the magenta regions in Fig~\ref{fig:PDtwospecieslikeone} are separated by the pale yellow line, where $\Lambda_{3,4}$ are purely imaginary: this purely oscillatory state is strongly affected by the nonlinearities in equations \eqref{eq:rho} and \eqref{eq:ptruncated}.

The modes $\Lambda_{3,4}$ possess a complex structure that is visible in the myriad effects that emerge at different lengthscales.  For finite $\kappa$ and vanishing $k$, we Taylor expand the non-conserving eigenmode as $\Lambda_{3} = -2D_r-k^2 D'+O(k^4)$, with a modified diffusion coefficient 
\begin{equation}\label{eq:diffPrime}
D' = \frac{\bar{\rho}\,\bm{\Omega}}{\kappa^2D_c}\cdot\left(\frac{v}{2D_r} \bm{\alpha}-\bm{\beta}\right)-\frac{v^2}{6D_r}+D.
\end{equation}
An instability at $k \to 0$ is ruled out by $D_r$, which is constrained to be positive. However, $D'$ can change sign signaling an instability at finite $k \approx \sqrt{|D_r/D'|}$. At low $k$, the conserved mode is $\Lambda_4 = -k^2 D'' +O(k^4)$, where the modified diffusion constant is
\begin{equation}\label{eq:diffDprime}
\begin{aligned}
D'' =  D^{\rm eff}(0)+\frac{\bar{\rho}\,\bm{\alpha}}{\kappa^2D_c}\cdot\left( \bm{\mu}-\frac{v}{2D_r}  \bm{\Omega}\right),
\end{aligned}
\end{equation}
and it can change sign leading to active phase separation when the effect of the phoretic interaction prevails over diffusion. $D''$ retains its positive sign if the combinations $\bm{\alpha} \cdot \bm{\mu}$ and $\bm{\alpha} \cdot \bm{\Omega}$  are positive and negative respectively - significant departures can cause $D''$ to flip sign and signal an instability. The first contribution in $D''$ arises from interactions between colloids when each acts as a point source of chemicals. The interactions are analogous to screened electrostatic ones, where the positive (respectively, negative) sign of $\mu_a$ determines whether the interactions are repulsive (attractive).
This scenario is in contrast with the expected phenomenology, where colloids interact via long-ranged chemical fields leading to gravitational collapse or electrostatic screening. ~\cite{PhysRevLett.112.068302}. This contribution can be understood as the screened analog of a Keller-Segel-like interaction, which was reported for the single species case in~\cite{PhysRevE.89.062316}. The second term is a combination of the collective turning of the polarization to point towards a local accumulation of the substrate and consequent drift in that direction due to self-propulsion; for single-species case, see~\cite{PhysRevE.89.062316} and appendix \ref{app:OneSpecies}. To summarize, at the smallest values of $k$, only the conserving mode can trigger instability.

Complex modes arise when the discriminant in equation~\eqref{eq:eigs2} becomes negative, i.e., $\Delta_2<0$. As $\Delta_2(0) = D_r^2$ is a positive quantity, it can change sign only at finite $k$, meaning that the response is oscillatory only at finite lengthscales. These \textit{exceptional} points, at which the complex eigenvalues emerge, are given by the roots of the cubic equation $\Delta_2(k^2) = 0$. They are denoted by the dashed line in the phase diagram in figure~\ref{fig:PDtwospecieslikeone}. As seen in the expression for $\Delta_2$ in equation~\eqref{eq:eigs2}, if $v$ is negligible, it is approximately a function of $k/\kappa$, and not just $k$. This means that the value of $k$ where the exceptional points appear scales with $\kappa$, approaching $k \to 0$ as $\kappa \to 0$, i.e. for long-range interactions. This effect can be visualized in figure \ref{fig:Eigtwo} (a), where we show how reducing the value of $\kappa$, the eigenvalues - in this case, associated with a stable disordered phase - become complex at smaller values of $k$. As $\kappa \to 0$, $\Delta_2$ approaches a finite value such that the model predicts (almost) global oscillations with frequency $\sqrt{|\Delta_2|}$, for example at $k=\kappa$ 
\begin{equation}
 \Delta_2(\kappa)
    = \bar{\rho}^2 \left( \frac{\bm{\alpha}\cdot\bm{\mu}-\bm{\beta}\cdot\bm{\Omega}}{4D_c}\right)^2 + D_r^2 +\frac{ \bar{\rho}}{2D_c}\left[v \bm{\alpha}\cdot\bm{\Omega}-D_r\left(\bm{\beta}\cdot\bm{\Omega}+\bm{\alpha}\cdot\bm{\mu}\right)\right] + O(\kappa^2).
\end{equation}
In this regime, complex eigenvalues emerge when the combination $(\bm{\beta} \cdot \bm{\Omega} + \bm{\alpha} \cdot \bm{\mu} )>0$, ensuring that $\Delta_2<0$. 
Even in the absence of chemotactic alignment ($\Omega_a=0$), we obtain an (unstable) oscillating phase for $\bm{\alpha}\cdot\bm{\mu}$ sufficiently negative. Being $\Lambda_4$ associated with the non-conserved mode in the system, we deduce that an interplay of two fields is essential for oscillations. 
The oscillations are spontaneous if the real part of $\Lambda_{3,4}$ is negative, which is given according to equation~\eqref{eq:eigs2} by 
\begin{equation}
\mbox{Re}(\Lambda_{3,4}) = -D_r -D k^2 + \frac{K(k) \bar{\rho}}{D_c} (\bm{\alpha}\cdot\bm{\mu} - \bm{\beta}\cdot\bm{\Omega}).
\end{equation}
The condition $\mbox{Re}(\Lambda_{3,4}) = 0$ - the green line in figure~\ref{fig:PDtwospecieslikeone} - corresponds to the threshold of a Hopf bifurcation. Orientational diffusion stabilizes the homogeneous disordered phase at $k=0$, hence spontaneous oscillations also occur only at finite $k$ and for $(\bm{\beta} \cdot \bm{\Omega} - \bm{\alpha} \cdot \bm{\mu} )<0$. We expect a regime that should be dominated by phase separation as well as aster formation.
If the contribution of $D$ and $v$ in $\Lambda_{3,4}$ are non-negligible, the eigenmodes $\Lambda_{3,4}$ depend both on $k$ and $\kappa$ creating the possibility of oscillatory instabilities at finite $k$ that are controlled by $v$ and the combination $\left(\bm{\alpha}\cdot\bm{\Omega}+{k^2}\bm{\beta}\cdot\bm{\mu}/3\right)$. In this regime of parameters, - as we showcase in figure \ref{fig:Eigtwo} (b) - as the screening parameter $\kappa$ decreases, oscillations may emerge at increasing values of $k$.

As a concluding remark, we mention that the eigenvalues $\Lambda_{3,4}$ are formally similar to the one-species case; we refer to appendix \ref{app:OneSpecies} for the details. This similarity becomes a qualitative equivalence whenever it is possible to factor out the scalar product of the vectorial production rates and phoretic interactions $\bm{\alpha}$, $\bm\beta$, $\bm\mu$ and $\bm\Omega$. For instance, if $\alpha_1=\alpha_2$ and $\beta_1=\beta_2$ we have $\bm\alpha\cdot\bm\mu=\alpha\mu_{\rm eff}$, $\bm\alpha\cdot\bm\Omega=\alpha\Omega_{\rm eff}$, $\bm\beta\cdot\bm\mu=\beta\mu_{\rm eff}$, and $\bm\beta\cdot\bm\Omega=\beta\Omega_{\rm eff}$ such that $\mu_{\rm eff}$ and $\Omega_{\rm eff}$ can be interpreted as one species effective phoretic couplings.

\begin{figure}
\centering
\includegraphics[width = \linewidth]{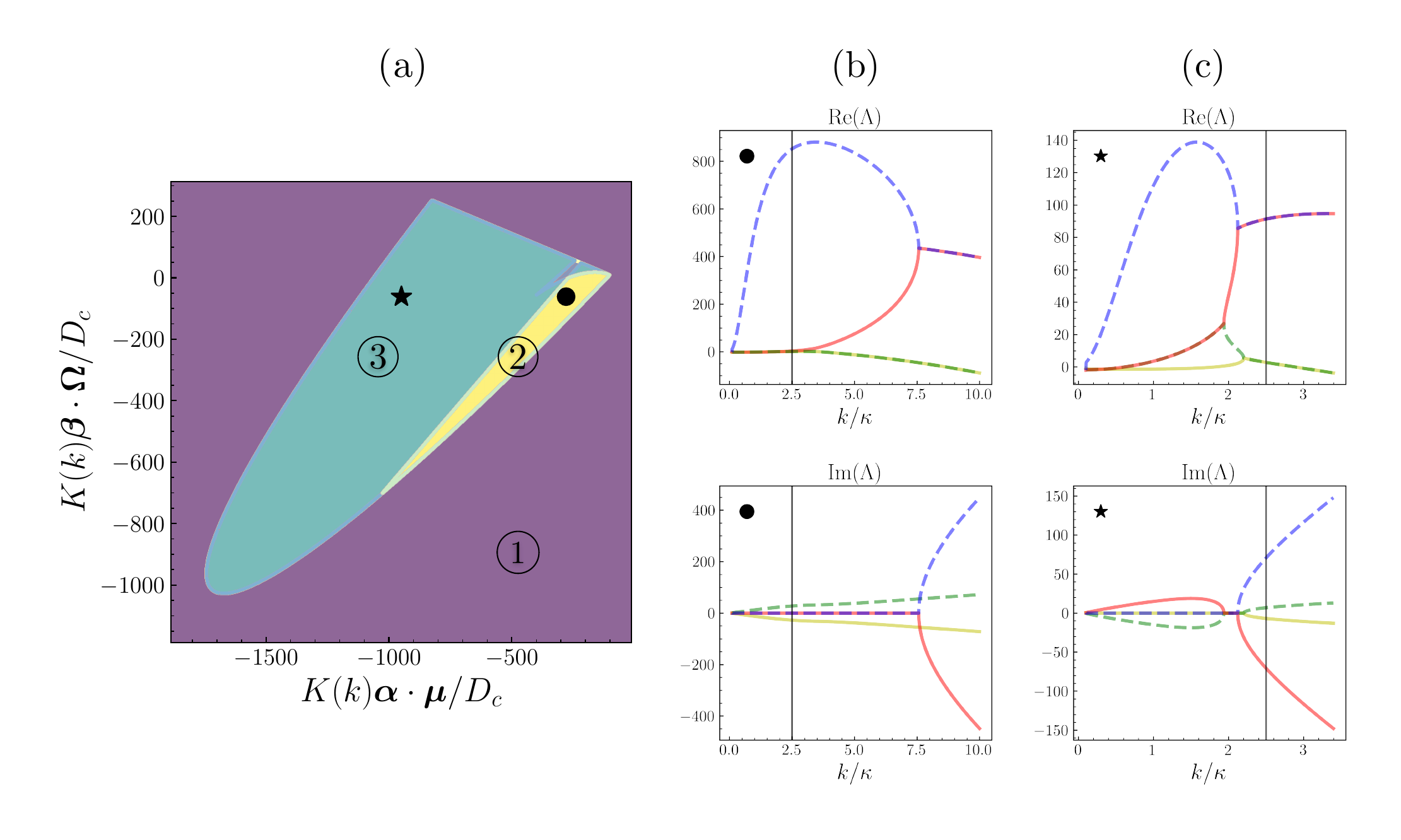}
\caption{In panel (a) we show the phase diagram for generic values of the non-active parameters $D$, $D_r$, and $v$. In this case, the system may display four unstable modes. For instance, in the diagram of panel (a): the yellow area \raisebox{.5pt}{\textcircled{\raisebox{-.9pt} {2}}} corresponds to two complex unstable and real unstable eigenvalues, while the green one \raisebox{.5pt}{\textcircled{\raisebox{-.9pt} {3}}} to four complex unstable eigenvalues. The violet area \raisebox{.5pt}{\textcircled{\raisebox{-.9pt} {1}}} indicates the region of the parameters where at least one eigenvalue is stable. The parameters are given by $D_1=D_2=1$, $D_{r,1}=1$, $D_{r,2}=1.2$,  $v_1=10$, $v_2=20$, and $k/\kappa =2.5$ . In panels (b) and (c) we represent the real and imaginary parts of the four eigenvalues of the dynamical matrix for the parameters corresponding to the bullet and star symbol in the phase diagram as a function of $k$. The vertical black line identifies the value of $k/\kappa$ of the phase diagram in the (a) panel.}\label{fig:4unstable}
\end{figure}

For the general case $D_{r,1}\neq D_{r,2}$, $D_1\neq D_2$ and $v_1\neq v_2$ all four eigenvalues may be unstable, and the picture becomes richer. We represent an instance of this behavior in figure \ref{fig:4unstable} (a), which displays the region of the phase diagram where eigenvalues are all unstable for a given value of $k$. 
To fix ideas, if we set $\alpha_a,\beta_a>0$, all the modes are unstable for sufficiently high and negative values of $\mu_a$ and $\Omega_a$: particles are expected to aggregate because of chemotaxis, while their orientation is expected to point away from the induced high-density region. These facts, together with diffusion, now occurring over different timescales for the two species, lead to a strongly unstable phase.
In figure \ref{fig:4unstable}(b)-(c), we plot the four eigenvalues of the dynamical matrix as a function of the wave vector $k$ for two specific choices of the parameter values of the phase diagram in figure \ref{fig:4unstable} (a).

To summarize the main points in this section, we find that phase separation driven by phoretic interactions emerges as the dominant behavior at the largest lengthscales. Oscillations, which can also be spontaneous, occur generically at finite wave numbers. At precisely $\kappa = 0$, both the conserving and the non-conserving modes can be unstable. In this regime, we truly have two hydrodynamic modes: one corresponding to number conservation, and the other an order parameter field that can undergo a phase transition and pick up a finite value.

\section{Adiabatic elimination of the polarization fields: coupled active densities}\label{sec:twospeciesrelaxed}
As discussed in the previous section, the dynamical properties of the two species of Janus particles are described by the conserved slow density mode $\rho_a(\bm{r},t)$ and the longitudinal part of the non-conserved polarization field ${p}_{\parallel,a}(\bm{r},t)$. Strictly speaking, $\delta p_{\parallel,a}$ is a slow variable at lengthscales that are small compared to the screening length $\kappa^{-1}$. While probing the dynamics at the largest lengthscales, and for finite $\kappa^{-1}$, we can assume that $\delta p_{\parallel,a}$ relaxes faster than $\delta \rho_a$ and express the former in terms of the latter. Equivalently, the field $\dwak$ in equation \eqref{eq:FTlineqtime} adapts instantaneously to the time variations of $\drak$ according to
\begin{equation}\label{eq:FTlineq}
\begin{aligned}
  \dwak&=-ik\sum_b\Gamma_{ab}(k)\drbk.
\end{aligned}
\end{equation}
The matrix $\Gamma$ appearing in equation \eqref{eq:FTlineq} is defined in terms of those appearing in equation~\eqref{eq:dynamicalmatrices} as $\Gamma(k)\equiv\left[\mathcal{G}^{rr}(k)\right]^{-1}\mathcal{G}^{rt}(k)$. Explicitly, its entries are given by 
\begin{equation}\label{eq:relaxcoeff}
\begin{aligned}
&\Gamma_{ab}(k)\equiv\frac{2}{3 \bar{D}_{a}(k)}\left(\frac{v_a}{2}\delta_{ab}-\bar{\rho}_a\Omega_a\Pi_b(k)\right),
\end{aligned}
\end{equation}
where $\bar{D}_a(k) = 2 D_{r,a} + D_a k^2$, and the redefined substrate production rate $\Pi_a(k)$ reads
\begin{equation}\label{eq:Pi}
\Pi_{a}(k)\equiv \frac{1}{\kappa^2D_c  + k^2\left[D_c-2\left(\bar{\rho}_1 {\beta_1  \Omega_1}/{\bar{D}_1}(k)+\bar{\rho}_2{\beta _2  \Omega _2}/{\bar{D}_2(k)}\right)/3\right]}\left(\alpha_a-k^2\frac{v_a\beta_a}{3 \bar{D}_a(k)}\right).
\end{equation}
The matrix $\Gamma$ quantifies the linear response matrix connecting the polar field to variations of the density, while $\Pi_a(k)$ is the effective rate of substrate production by species $a$.
\begin{figure}
\centering
\includegraphics[width = 0.5\linewidth]{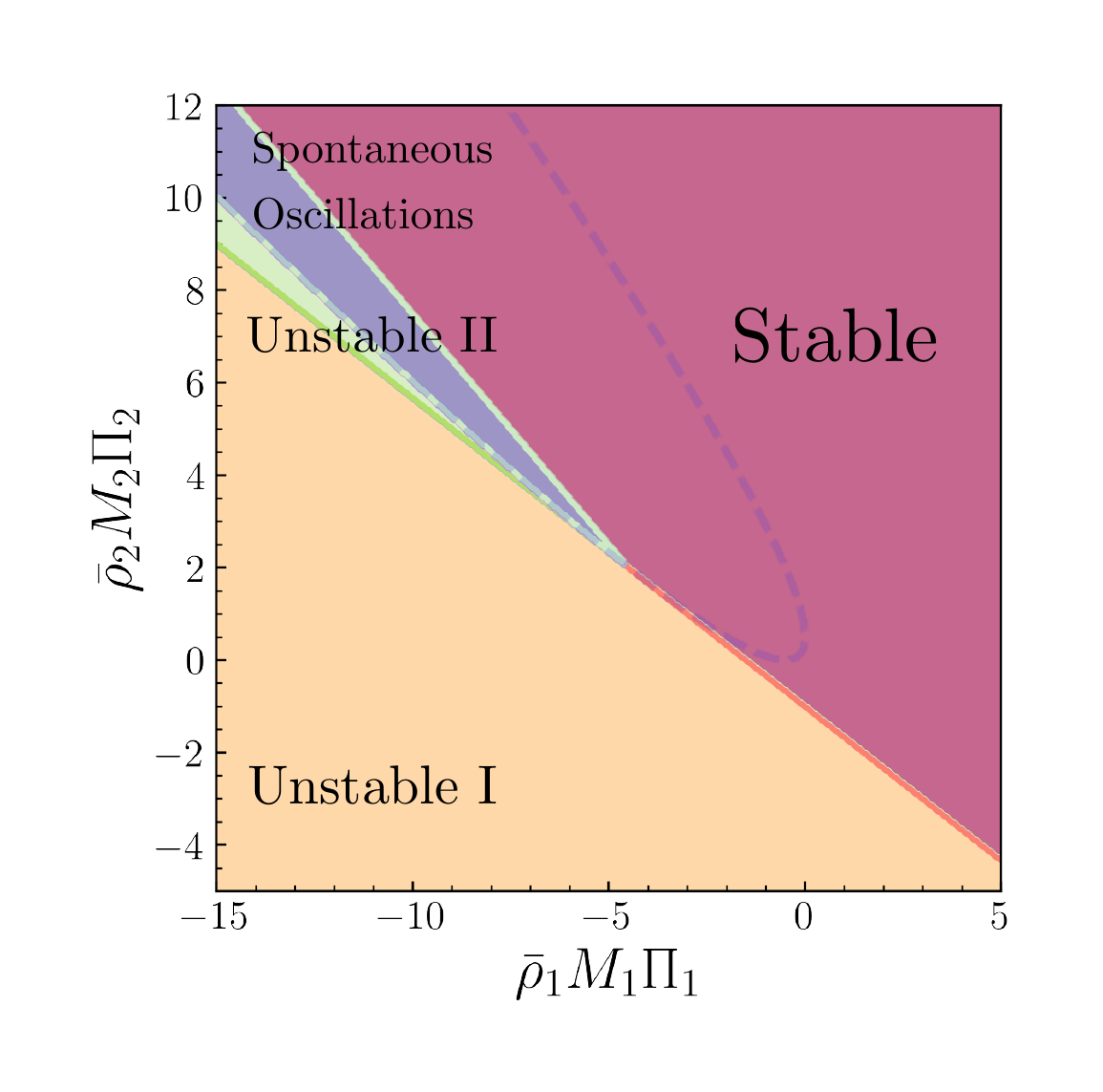}
\caption{Phase diagram from the linear stability analysis. The magenta area identifies the values of $\bar{\rho}_1 M_1\Pi_1$ and $\bar{\rho}_2 M_2\Pi_2$ corresponding respectively to a linear stable regime. The ``Unstable I" (orange) region corresponds to the case where the system presents one unstable and one stable mode. In the ``Unstable II'' (green) one, the modes are still real but both unstable. In the correspondence of the blue region of the phase diagram, ``Spontaneous Oscillations'' arise in the system, meaning the two eigenvalues are complex conjugate with positive real part (unstable).   
The parabola separates the parameter space between real and complex eigenvalues.}\label{fig:phasediagram_dim}
\end{figure}
We now investigate the behavior of the system in this regime, where the only relevant fields are the densities of the two species of particles. 
A necessary condition for this approximation to hold is to have positive eigenvalues of $\mathcal{G}^{rr}(k)$, thus ensuring the relaxation of the polar field in equation \eqref{eq:FTlineqtime}. The enslaving fails when the determinant of $\mathcal{G}^{rr}$ vanishes,  which happens when one or both the eigenvalues of $\mathcal{G}^{rr}$ are null.  Large enough $D_{r,a}$ rules out this possibility, ensuring the validity of the enslaving.
As discussed in the previous section, the approximation holds also for sufficiently large $\kappa^2$, ruling out aster condensation. The denominator of $\Pi_a(k)$ is proportional to the determinant of $\mathcal{G}^{rr}$, which is positive whenever the relaxation approximation holds. Therefore, the sign of $\Pi_a(k)$ follows from those of $\alpha_a$, $\beta_a$, the wave vector $k$, and their relative amplitude. Note that in the case of uniform production or consumption of the chemical $\beta_a=0$, the fast relaxation approximation is well defined and the effective production rate reduces to $\Pi_a(k)=\alpha_a/[D_c(k^2+\kappa^2)]$, whose sign depends only on that of $\alpha_a$. Substituting the expression of the polar field in equation \eqref{eq:FTlineq} in the equation for $\drak$ in equation \eqref{eq:FTlineqtime}, one finds 
\begin{equation}\label{eq:rholinear}
\begin{aligned}
\partial_t\drak=k^2\sum_{b}\mathcal{G}_{ab}(k)\,\drbk,
\end{aligned}
\end{equation}
 where we have introduced the effective diffusion matrix $\mathcal{G}(k)\equiv\mathcal{G}^{tt}(k)+\mathcal{G}^{tr}(k)\,\Gamma(k)$. One can show that the entries of $\mathcal{G}(k)$ can be expressed as
 \begin{equation}\label{eq:Dk}
     \mathcal{G}_{ab}(k)=-\delta_{ab}D^{\rm eff}_a(k)-\bar{\rho}_a M_a(k)\Pi_b(k),
 \end{equation}
where, we define an effective mobility $M_a$ and an effective diffusivity $D_a^{\rm eff}$ as 
 \begin{equation}
    M_{a}(k)\equiv \mu_a-\frac{2}{3}\frac{v_a\Omega_a}{\bar{D}_a(k)},\quad D_a^{\rm eff}(k)=D_a+\frac{v_a^2}{3\bar{D}_a(k)}.
\end{equation}
Interestingly enough, the matrix $\mathcal{G}(k)$ inherits the same factorized structure as its building blocks in equations \eqref{eq:dynamicalmatrices}, i.e., a diagonal contribution coming from diffusion and self-propulsion, and contributions are multiples of phoretic mobility and chemical production rate as in equation \eqref{eq:structure}. Note that, while $D_a^{\rm eff}$ is always positive, the sign of the effective mobility $M_a$ depends on the relative amplitudes of the phoretic drift and alignment interactions. 
By splitting the symmetric and antisymmetric parts of the dynamical matrix $\mathcal{G}$ as
\begin{equation}\label{eq:linMatrixNRCH}
\mathcal{G}=-\left(
\begin{array}{cc}
 D_1^{\rm eff}+\bar{\rho}_1 M_1\Pi_1 & \bar{\rho}_1 M_1\Pi_2 \\[5pt]
    \bar{\rho}_2 M_2\Pi_1 & D_2^{\rm eff}+\bar{\rho}_2 M_2\Pi_2\\
\end{array}
\right)=\left(
\begin{array}{cc}
 d_1 & \chi+\psi \\[5pt]
    \chi-\psi & d_2\\
\end{array}
\right),
\end{equation}
one can identify self-interaction species term  $d_a = \mathcal{G}_{aa}$, the symmetric (reciprocal) contribution to the interaction between the two different species $\chi = (\mathcal{G}_{12}+\mathcal{G}_{21})/2$, and its anti-symmetric (nonreciprocal) one $\psi = (\mathcal{G}_{12}-\mathcal{G}_{21})/2$. By its definition, $\chi$ is symmetric under the exchange of the coefficients of the two species, while $\psi$ reverses sign under the same transformation. At leading order of a small-$k$ expansion the non-reciprocity parameter $\psi(k)$ reads

\begin{equation}\label{eq:NRCHparameters}
\begin{aligned}
    \psi(k)=&\,\psi_0+k^2\psi_2+O(k^2),\\
    \psi_0 =& -\frac{1}{2\kappa^2D_c}\left[\left( \bar{\rho}_1 \mu_1 \alpha_2-\bar{\rho}_2 \mu_2 \alpha_1 \right) -  \left( \bar{\rho}_1 \alpha_2 \frac{v_1\Omega_1}{3D_{r,1}}  - \bar{\rho}_2 \alpha_1  \frac{v_2\Omega_2}{3D_{r,2}} \right) \right],\\
    \psi_2 =&-\frac{1}{12\kappa^2 D_c}\left[\frac{\bar{\rho}_1v_1\Omega_1D_1\alpha_2}{ (D_{r,1})^2}-\frac{\bar{\rho}_2v_2\Omega_2D_2\alpha_1}{ (D_{r,2})^2}
    -\left( \bar{\rho}_1 \mu_1 \frac{v_2\beta_2}{D_{r,2}}  - 
    \bar{\rho}_2 \mu_2 \frac{v_1\beta_1}{D_{r,1}} \right)
    + \frac{v_1 v_2 }{3D_{r,1} D_{r_2}} \left( \bar{\rho}_1    \Omega_1 \beta_2 - \bar{\rho}_2    \Omega_2 \beta_1 \right)\right]\\
    &-\frac{\psi_0}{\kappa^2D_c}\left(D_c-\bar{\rho}_1\frac{\beta_1  \Omega_1}{3D_{r,1}}-\bar{\rho}_2\frac{\beta _2  \Omega _2}{3 D_{r,2}}\right),
\end{aligned}
\end{equation}
where the first term identifies the macroscopic effective nonreciprocal interaction between the two different species. The parameter $\psi_0$ has been introduced in~\cite{NRCH,you2020nonreciprocity} in a minimal model for nonreciprocal interaction between multiple species, called the nonreciprocal Cahn Hilliard (NRCH) model. The introduction of this term in a phase-separating system of many conserved densities leads to arrested phase separation, broken spatial parity, and broken time-reversal symmetry,  producing traveling waves and patterns. The second term couples the two species at the fourth order in gradients and can be interpreted as a nonreciprocal surface tension \cite{PhysRevLett.131.148301}. Interestingly enough, for $\Omega_a = \beta_a = 0$, the expression of $\psi$ simplifies to
\begin{equation}\label{eq:psiSimple}
\psi = -\frac{\bar{\rho}_1 \mu_1 \alpha_2-\bar{\rho}_2 \mu_2 \alpha_1}{2D_c(\kappa^2 + k^2)},
\end{equation}
indicating that uniformly coated colloids interacting via a screened chemical field generates nonreciprocal couplings and all orders in gradients, the two nonreciprocal at lowest order in gradients are related simply as $\psi_2 = - \psi_0 \kappa^{-2} $. Non-zero $\beta_a$ and $\Omega_a$ take us away from the simple relation between $\psi_{0}$ and $\psi_2$, which can now be tuned independently of one another. 
 
The self interaction $d_a$ and the reciprocal interaction $\chi$ can be expanded similarly as $d_a = d_{a,0} + d_{a,2} k^2$, and $\chi = \chi_0 + \chi_2 k^2$. The coefficient $d_{a,0}$ is the strength of self-interaction for species $a$, while $\chi_0$ is the effective reciprocal interaction. The terms occurring at higher orders in gradients, namely $d_{2,a}$, and $\chi_{2}$ are the coefficients for interfacial tension. Recall that the eigenmodes in the previous section had contributions from symmetric combinations of the phoretic parameters which could be written as dot products such as $\bm{\alpha}\cdot \bm{\mu}$. Similar simple relations hold for the elements of $\mathcal{G}$. We can express $d_a$, $\psi$, and $\chi$ compactly by introducing the following matrices
\begin{equation}\label{eq:I2}
\bm{\sigma}_1 = 
\begin{pmatrix} 0  & 1 \\ -1 & 0 \end{pmatrix},\;\;\; 
\bm{\sigma}_2 = 
\begin{pmatrix} 0  & 1 \\ 1 & 0 \end{pmatrix},\;\;\;
\bm{\sigma}_3 = 
\begin{pmatrix} 1  & 0 \\ 0 & -1 \end{pmatrix},
\end{equation}
and the vectors of parameters

\begin{equation}
    \omOne = \left(\frac{v_1 \bar{\rho}_1 \Omega_1}{\bar{\rho}D_{r,1}}, \frac{v_2 \bar{\rho}_2 \Omega_2}{\bar{\rho}D_{r,2}}\right),\;\;\;\;\bm{\beta}' = \left(\frac{v_1 \beta_1}{D_{r,1}}, \frac{v_2\beta_2}{D_{r,2}}\right),\;\;\;\;\omTwo = \left(\frac{\bar{\rho}_1 D_1 \Omega_1}{\bar{\rho}(D_{r,1})^2}, \frac{\bar{\rho}_2 D_2 \Omega_2}{\bar{\rho}(D_{r,2})^2}\right),\;\;\text{and}\;\;\bm{\Omega}'''=\left(\frac{\bar{\rho}_1 \Omega_1}{\bar{\rho}D_{r,1}}, \frac{\bar{\rho}_2 \Omega_2}{\bar{\rho}D_{r,2}}\right),
\end{equation} 
and the auxiliary scalar functions
\begin{equation}
A_0(\mathcal{M}) = -\frac{\bar{\rho}}{2\kappa^2D_c} \left(  \bm{\mu} \cdot \mathcal{M} \cdot \bm{\alpha} - \frac{1}{3} \omOne \cdot \mathcal{M} \cdot \bm{\alpha}   \right)=-\frac{\bar{\rho}}{2\kappa^2D_c} \left(  \bm{\mu} - \frac{1}{3} \omOne    \right)\cdot \mathcal{M} \cdot \bm{\alpha},
\end{equation}
\begin{equation}
A_2(\mathcal{M}) = -\frac{\bar{\rho}}{12\kappa^2 D_c}  \left( \omTwo \cdot \mathcal{M} \cdot \bm{\alpha}
    - \bm{\mu} \cdot \mathcal{M} \cdot \bm{\beta}' 
    + \frac{v_1 v_2 }{D_{r,1} D_{r_2}} \bm{\Omega} \cdot \mathcal{M} \cdot  \bm{\beta} \right) -\frac{A_0(\mathcal{M})}{\kappa^2D_c}\left(D_c-\bar{\rho}\bm{\beta}\cdot\bm{\Omega}'''\right).
\end{equation}
Finally, we can write the coefficients of the model as follows

\begin{equation}\label{eq:Elements}
\begin{split}
    \psi_0 &= A_0(\bm{\sigma}_1),\\
    \chi_0 &= A_0(\bm{\sigma}_2),\\
    d_{1,0} &= -D_1^{\rm{eff}}(0)+ A_0(\mathbb{I} + \bm{\sigma}_3),\\
    d_{2,0} &= -D_2^{\rm{eff}}(0) +A_0(\mathbb{I} - \bm{\sigma}_3),
\end{split}
\quad\quad
\begin{split}
    \psi_2 &= A_2(\bm{\sigma}_1), \\
    \chi_2 &= A_2(\bm{\sigma}_2), \\
    d_{1,2} &= -D_1v_1^2/[12(D_{r,1})^2]+A_2(\mathbb{I} + \bm{\sigma}_3), \\
    d_{2,2} &= -D_2v_2^2/[12(D_{r,2})^2]+A_2(\mathbb{I} - \bm{\sigma}_3).
\end{split}
\end{equation}
In equation \eqref{eq:Elements}, $\psi_{0,2}$ are pseudoscalars while the rest are scalar in the exchange of the two species. 
\subsection{Stability diagram} The linear stability of the homogeneous state is determined by the eigenvalues $\Lambda_{1,2}(k)$ of $\mathcal{G}(k)$ which can be expressed compactly in terms of the effective mobility $\bar{\rho}_a M_a$ and the production rate $\Pi_a(k)$ as
\begin{equation}\label{eq:eigenvalues}
\begin{aligned}
\Lambda_{1,2}
&=-\left(\frac{D_1^{\rm eff}+D_2^{\rm eff}+\bar{\rho}_1 M_1\Pi_1+\bar{\rho}_2 M_2\Pi_2}{2} \pm \sqrt{\Delta}\right)\\
\Delta&=\left(\frac{D_1^{\rm eff}+\bar{\rho}_1 M_1\Pi_1-D_2^{\rm eff}-\bar{\rho}_2 M_2\Pi_2}{2}\right)^2+\bar{\rho}_1 M_1\Pi_1 \bar{\rho}_2 M_2\Pi_2.\\
\end{aligned}
\end{equation}
The growth rates of the eigenmodes are then given by $k^2\Lambda_{1,2}(k)$ which vanish as $k \to 0$, as a result of number conservation. For purely reciprocal interactions corresponding to $\psi=0$, $\mathcal{G}$ is symmetric and the eigenvalues are always real. Another case where the eigenvalues are always real is for equal effective diffusivities $D_1^{\rm eff}=D_2^{\rm eff}=D^{\rm eff}$. In this case, $\mathcal{G}$ is a rank one tensor of the form 
\begin{equation}
\mathcal{G} = -D^{\rm eff} \mathbb{I}  - (\bar{\rho}_1 M_1,\bar{\rho}_2 M_2)^{\rm T}(\Pi_1,\Pi_2). 
\end{equation}
The eigenmodes are $\Lambda_1=-D^{\rm eff}$,  $\Lambda_2=-(D^{\rm eff}+ \Trace{\mathcal{G}}) = -(D^{\rm eff} + \bar{\rho}_1 M_1 \Pi_1 + \bar{\rho}_2 M_2 \Pi_2)$. $\Lambda_1(k)<0$ is the stable mode, while $\Lambda_2(k)$ can be positive and trigger an instability for $D^{\rm eff} > -(\bar{\rho}_1 M_1\Pi_1+\bar{\rho}_2 M_2\Pi_2)$. For $D^{\rm eff}_1 \neq D^{\rm eff}_2$, and considering equation \eqref{eq:eigenvalues}, it is clear that complex eigenvalues can occur only if
\begin{equation}
 \chi^2-\psi^2 = \bar{\rho}_1 M_1\Pi_1 \bar{\rho}_2 M_2 \Pi_2 <0,
\end{equation}
whenever the nonreciprocal coupling exceeds its reciprocal counterpart; this condition is sufficient for $d_1=d_2$. Henceforth, without any loss of generality, we always assume that $\delta D = D_1 - D_2>0$. 

Similarly to section \ref{sec:twospecies}, we first present the phase diagram in the plane of parameters $\bar{\rho}_a M_a \Pi_a$. In figure~\ref{fig:phasediagram_dim}, the region where the eigenvalues are complex lies on the convex side of the parabola. The homogeneous state is stable in the magenta part. Both $\Lambda_{1,2}$ are unstable in the green region and signal an instability leading to phase separation. In the orange region, one of the two eigenvalues leads to an instability. In the purple region, the two eigenvalues are complex and unstable leading to oscillating densities. The phase diagram has the same topology for all values of $k$.   

 We will now discuss how the instabilities appear at different lengthscales, i.e. as $k$ is varied. Alternatively, if one keeps all the other parameters fixed and considers the eigenvalues as a function of $k$ only, the system state can pass through different phases, described as a \textit{state curve} parameterized by $k$. Next, we will discuss the instabilities that occur as $k \to 0$ and at finite $k$ and illustrate them on the phase diagram in figure \ref{fig:phasediagram_dim} using state curves.

\begin{figure}
\centering
\includegraphics[width = \linewidth]{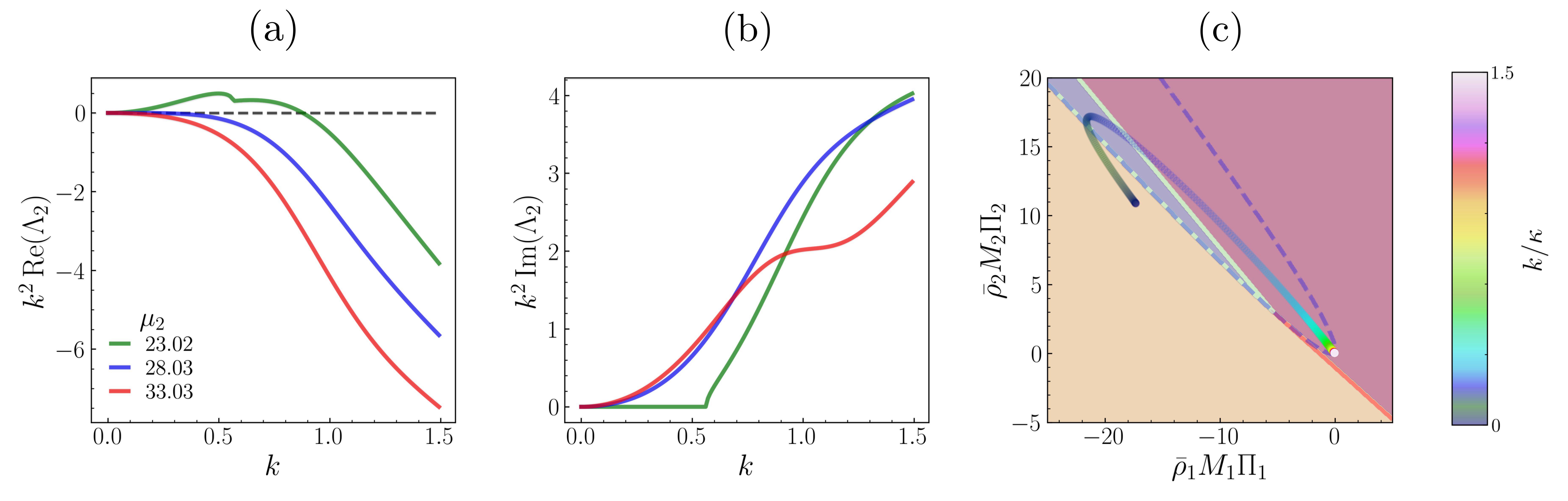}\caption{Type II of instability. Another type of instability that may occur in our conserved system can be attained whenever the eigenvalues are already unstable at small values of $k$, i.e., Re$(\Lambda_2(0))>0$. In (a) and (b) we show, respectively, the real and imaginary parts of the most unstable eigenvalue $\Lambda_2$ at the onset of the instability, which is triggered by variations of the parameter $\mu_2$.  In (c) the equivalent parametric plot of the state curve described by the system in the phase diagram is showcased: the initial point of the trajectory lies in the instability region and eventually ends in the stable one. The parameters are given by: $D_c=1$, $D_1=2$, $D_2=1.5$, $D_{r,1}=1$, $D_{r,2}=2$, $v_1=1$, $v_2=1.5$, $\bar\rho_1=\bar\rho_2=1$, $\alpha_1=4$, $\alpha_2=0.5$, $\beta_1=0.5$, $\beta_2=1$, $\mu_1=-1$, $\Omega_1=10$, $\Omega_2=5$. As $k$ varies, the phase diagram slightly changes quantitatively but not qualitatively, thus we draw it for a fixed value of $k$. In this figure, it has been chosen $k/\kappa=10$.}
\label{fig:PD_typeII}
\end{figure}

\subsection{Instabilities at vanishing $k$}
We expand $\Lambda_{1,2}$ in a Taylor series to obtain the following general expression at zeroth order in $k$ 

\begin{equation}\label{eq:LambdaZeroK}    \Lambda_{1,2}(0)=
\begin{cases}
    \frac{\Trace\mathcal{G}(0)}{2} \mp \sqrt{\Delta(0)}&\text{for }\Delta(0)>0\\
    \frac{\Trace\mathcal{G}(0)}{2} \mp i \sqrt{-\Delta(0)}&\text{for }\Delta(0)<0
\end{cases},
\end{equation}
where $\Delta(k) \equiv (\Trace \mathcal{G}/2)^2-\Det\mathcal{G}$. The coefficients in equation~\eqref{eq:LambdaZeroK} are
\begin{equation}
\Trace\mathcal{G}(0)=-\frac{1}{2}\left[D_1^{\rm eff}(0) +D_2^{\rm eff}(0) + \frac{\bar{\rho}\,\bm{\alpha}  }{D_c \kappa^2}\cdot \left(\bm{\mu} -  \bm{\Omega} \right)\right],
\end{equation}
and
\begin{equation}
\begin{aligned}
\Delta(0)=  \left[ \Trace\mathcal{G}(0) \right]^2 +\left(\mu_1-\frac{v_1\Omega_1}{3D_{r,1}}\right)\left(\mu_2-\frac{v_2\Omega_2}{3D_{r,2}}\right)\frac{\bar{\rho}_1\alpha_1\bar{\rho}_2\alpha_2}{(D_c\kappa^2)^2}.
\end{aligned}
\end{equation}
A system-wide instability arises whenever Re$(\Lambda_2)$ is positive, resulting in bulk phase separation of the two species. At large $k$, the eigenvalues are stable and real, ensuring the system's stability at small length scales.  
Accordingly, Re$\Lambda_2$ attains its maximum value at $k_c$. As the instability is turned off, $k_c\rightarrow 0$ and the value of the associated lengthscale $k_c^{-1}$ diverges since the maximum of the most unstable eigenvalue vanishes, i.e.,  Re$\Lambda_2(k_c\rightarrow 0)\rightarrow 0$. This corresponds to a type II of instability according to the Cross and Hohenberg classification \cite{RevModPhys.65.851}. If at the onset of the instability $\Lambda_2$ is real, it is further classified as \textit{stationary}, while in its complex counterpart as \textit{oscillatory}. We show an instance of this type of instability in figure \ref{fig:PD_typeII}. In panels (a) and (b) we showcase the real and imaginary parts of the most unstable mode $\Lambda_2$, and how instability is triggered while varying the $\mu_2$ parameter. In this specific case, we have an instance of stationary instability.
However, being $\Delta(0)$ a nonlinear combination of the phoretic parameters, an oscillatory instability develops for a large part of the parameter space, e.g.,  for $\mu_a, \Omega_a>0$, $\Delta(0)<0$ if 
\begin{equation}\label{eq:DeltaNeg}
\sign(\mu_1 - v_1 \Omega_1/3 D_{r,1})  \neq \sign(\mu_2 - v_2 \Omega_2/3 D_{r,2}),
\end{equation}
provided that both species produce chemicals, i.e., $\alpha_a>0$.
The associated system-wide oscillations occur with an angular frequency $\sqrt{|\Delta(0)|}$.
Another way to visualize the onset of the instability is displayed in figure \ref{fig:PD_typeII} (c), where we show how, by varying $k$, the parameters $\bar{\rho}_a M_a(k)\Pi_a(k)$ describe a state curve in the phase diagram, corresponding to the green curve in \ref{fig:PD_typeII} (a). At $k=0$ the system is the unstable region of parameters (orange area), and as the value of $k$ increases it crosses the complex unstable region (blue area), then the complex stable region (magenta area). Note that, as a consequence of diffusion, at large $k$ the system is always stable, as it can be seen from $\lim_{k\to\infty}\bar{\rho}_aM_a(k)\Pi_a(k)=0.$
Note that as the value of $k$ is varied, also the phase diagram changes. However, these changes are barely perceivable compared to the $\bar{\rho}_aM_a\Pi_a$ ones.  This fact allows us to give a meaningful qualitative representation of the $k$-parametrized state curve while drawing the diagram at a fixed value of $k$.
 
\subsection{Finite wavelength instabilities}
Instabilities in the system may also occur for Re$(\Lambda_2(0))<0$, i.e., the species are not separated at the macroscopic scale but produce patterns with a specific lengthscale. To characterize this type of instability, it is necessary to expand $\Lambda_{1,2}$ up to $k^2$ 
\begin{equation}\label{eq:ReII}
    \text{Re}(\Lambda_{1,2})=
    \begin{cases}
    \frac{\Trace \mathcal{G}(0)}{2}\mp \sqrt{\Delta(0)}+\frac{k^2}{4}\left[(\Trace \mathcal{G})''(0)\mp\frac{\Delta''(0)}{\sqrt{\Delta(0)}}\right]+O(k^4) &\text{for }\Delta(0)>0\\
    \frac{\Trace \mathcal{G}(0)}{2}+\frac{k^2}{4}(\Trace\mathcal{G})''(0)+O(k^4) &\text{for }\Delta(0)<0
    \end{cases},
\end{equation}
where $'$ denotes the derivative with respect to $k$. The expressions for terms that contribute at quadratic order in $k$ are given by 

\begin{equation}
\begin{aligned}
&(\Trace\mathcal{G})''(0)=-2\left\{ -\frac{D_1v_1^2}{12(D_{r,1})^2}-\frac{D_2v_2^2}{12(D_{r,2})^2} + \frac{\bar{\rho} \,\bm{\alpha}}{D_c \kappa^2}\cdot\left[ \frac{ \bm{\Omega}''}{6}  - \frac{1}{\kappa^2} \left(\bm{\mu}-\frac{\bm{\Omega}'}{3}\right)\right] \right\}, \\
&\Delta''(0)=\left[D_1^{\rm eff}(0) -D_2^{\rm eff}(0) + \frac{\bar{\rho}}{D_c \kappa^2} \left( \bm{\mu} -  \frac{\bm{\Omega}'}{3} \right) \cdot \bm{\sigma}_3 \cdot \bm{\alpha} \right] \left\{ -\frac{D_1v_1^2}{12(D_{r,1})^2}-\frac{D_2v_2^2}{12(D_{r,2})^2} + \frac{ \bar{\rho}\,\bm{\alpha}}{D_c \kappa^2}\cdot\left[ \frac{ \bm{\Omega}''}{6}  - \frac{1}{\kappa^2} \left(\bm{\mu}-\frac{\bm{\Omega}'}{3}\right)\right] \right\}\\
&\;\;\;\;\;\;\;\;\;\;+2\,\frac{\rho_1\alpha_1\rho_2\alpha_2}{(D_c\kappa^2)^2}\left\{ \left(\mu_1-\frac{v_1\Omega_1}{3D_{r,1}}\right)\frac{v_2\Omega_2D_2}{6(D_{r,2})^2}+ \left(\mu_2-\frac{v_2\Omega_2}{3D_{r,2}}\right)\frac{v_1\Omega_1D_1}{6(D_{r,1})^2}-\frac{2}{\kappa^2}\left(\mu_1-\frac{v_1\Omega_1}{3D_{r,1}}\right)\left(\mu_2-\frac{v_2\Omega_2}{3D_{r,2}}\right)\right\}.
\end{aligned}
\end{equation}
Note that the eigenmodes are invariant if the species indices $1$ and $2$ are swapped. $\Delta''(0)$ is invariant as it is a product of two quantities both of which reverse sign when under the swap $1 \longleftrightarrow 2$.
The system is stabilized at the shortest lengths by diffusive processes. A finite wavelength instability is triggered in the system if the most unstable mode Re$(\Lambda_2)$ becomes null at $k=k_-$ and then acquires a positive value. It reverses its sign again at $k=k_+$.  In this case, there exists a wave number $k_c$ intermediate between $k_{-}$ and $k_+$ where Re$(\Lambda_2)$ attains its maximal value, physically associated with pattern formation at the lengthscale $k_c^{-1}$. An approximate expression for $k_-$ can be retrieved from equation \eqref{eq:ReII} as 
\begin{equation}
    k_-=
    \begin{cases}
     2\sqrt{-\frac{\Trace\mathcal{G}(0)/2-\sqrt{\Delta(0)}}{(\Trace\mathcal{G})''(0)-\Delta''(0)/\sqrt{\Delta}(0)}}& \text{for }\Delta(0)>0\\[10pt]
     \sqrt{-2\frac{\Trace\mathcal{G}(0)}{(\Trace\mathcal{G})''(0)}}& \text{for }\Delta(0)<0,
    \end{cases}
\end{equation}
given that the square root exists.

\begin{figure}
\centering
\includegraphics[width = \linewidth]{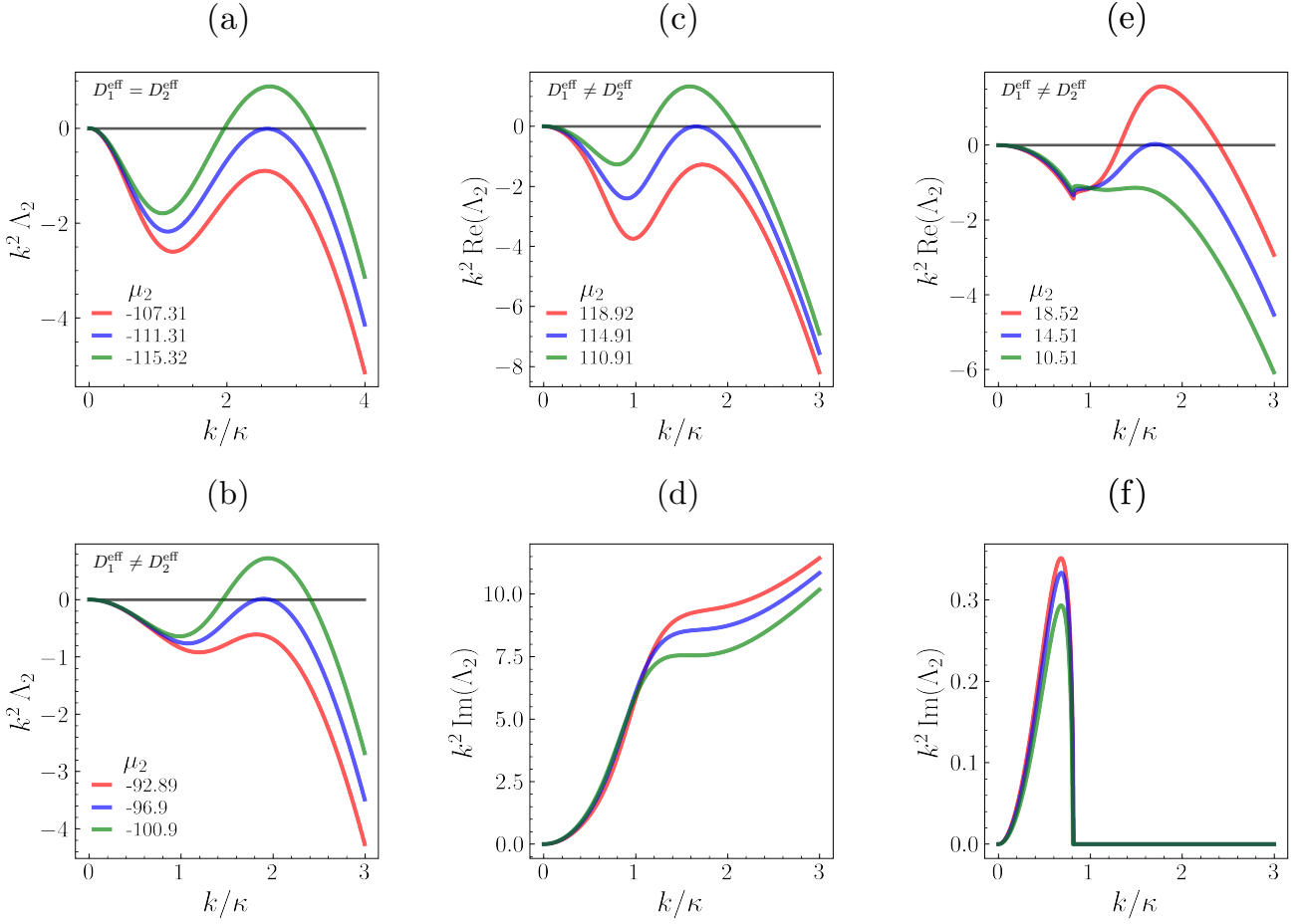}
\caption{We represent different instances of the onset of Turing instability, upon variation of the $\mu_2$ parameter. Whenever an instability arises it is characterized by the real part of the most unstable mode Re$(\Lambda_2)$ becoming positive. In panel (a), we consider the case where $D_1^{\rm eff}=D_2^{\rm eff}$: the eigenvalues are both real and instability may occur only because of the mode associated with $\Lambda_2$. The parameters are given by: $D_c=D_1=D_2=\bar\rho_1=\bar\rho_2=v_1=v_2=1$, $D_{r,1}=D_{r,2}=5$, $\alpha_1=1.5$, $\alpha_2=0.1$, $\beta_1=1$, $\beta_2=-0.5$, $\mu_1=10$, $\Omega_1=5$, $\Omega_2=-10$. In (b), the two effective diffusivities are different and the eigenvalues are still real. The parameters in panel (b) differing from those in (a) are: 
$D_1=2$, $D_2=1.5$, $D_{r,1}=1$, $D_{r,2}=2$, $v_2=1.5$. 
In panels (c) and (d) we represent respectively the real and imaginary parts of $\Lambda_2$ in the case where for all values of $k$ the eigenvalues are complex conjugate, corresponding to an oscillatory Turing instability. The parameters differing from (b) are:  $\alpha_1=4$, $\alpha_2=0.5$, $\beta_1=0.5$, $\beta_2=1$, $\mu_1=-10$, $\Omega_1=10$, $\Omega_2=5$.
Similarly, (e) and (f) correspond to the case where the system presents stable complex eigenvalues at small values of $k$, while the instability is associated with a stationary pattern. The parameters from (c) and (d) are $\alpha_1=2.5$, $\alpha_2=0.1$, $\beta_1=0.4$, $\Omega_1=5$, $\Omega_2=10$.
}\label{fig:turing}
\end{figure}

\begin{figure}
\centering
\includegraphics[width = \linewidth]{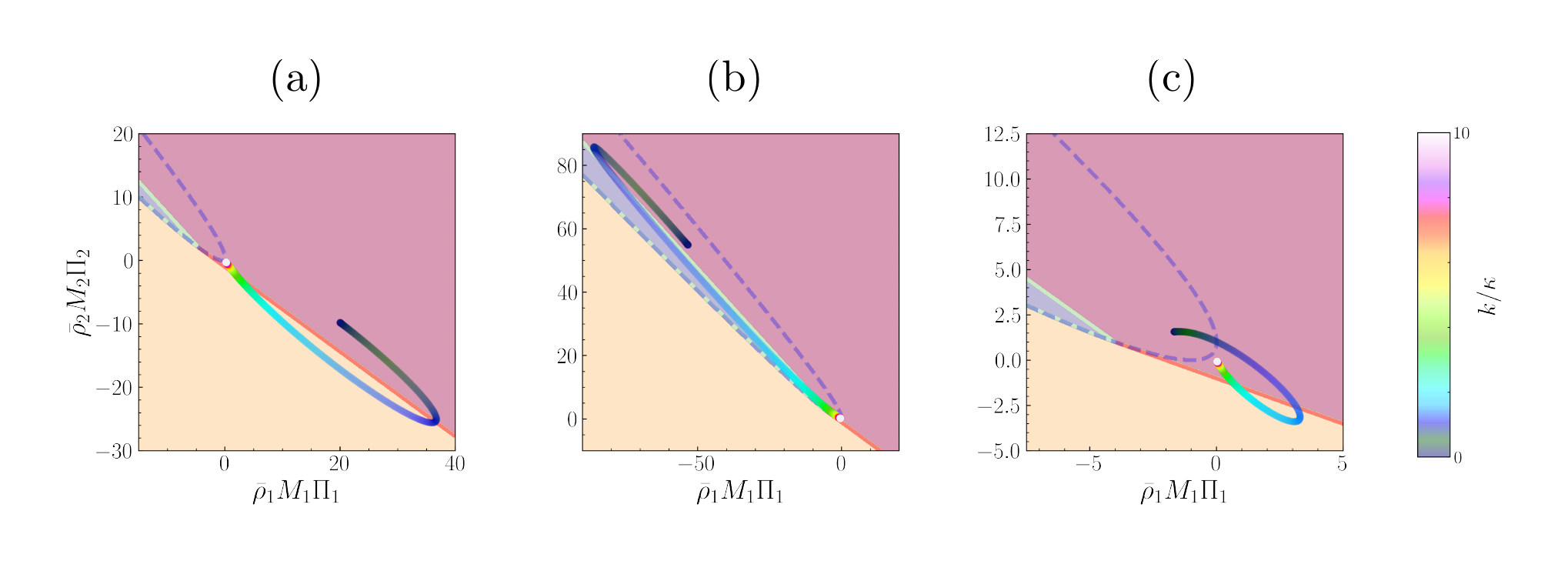}
\caption{Phase diagram representation of the Turing instability. These three figures show how, by keeping all system parameters fixed, $k^2$Re$(\Lambda_2)$ describes a state curve as $k$ is varied in the phase diagram. Similarly to figure \ref{fig:phasediagram_dim}, the magenta area corresponds to the stable region of the homogeneous disordered state, while blue and orange refer respectively to complex and real unstable eigenvalues. The parametric curves correspond in panels in (a), (b), and (c) respectively to the unstable (green) ones in figures \ref{fig:turing} (b), \ref{fig:turing} (c)-(d), and
\ref{fig:turing} (e)-(f). All three curves start at $k=0$ on the stable side of the phase diagram, and they end up in the origin, stable point of the phase diagram, as $\bar\rho_aM_a\Pi_a\rightarrow 0$ as $k\rightarrow \infty$, while they cross to the instability region at intermediate values of $k$. This is the benchmark of the Turing instability. As in figure \ref{fig:PD_typeII}, the phase diagram changes very little as $k$ is changed, and for simplicity, we have set $k/\kappa= 10$.}\label{fig:PD_typeI}
\end{figure}

At arbitrary $k$, this type of behavior is well represented in figure \ref{fig:turing}, where we provide a few examples of the onset of the instability. Note that if the eigenvalues are real at $k=k_- \text{ (respectively, }k_+ )$ then $\Lambda_2=0$ and $\Lambda_1\le 0 (\ge 0)$, and the state curve crosses the line corresponding to $ \Det\mathcal{G}=0$ at $k_- \text{ (respectively, }k_+ )$ from its stable (unstable) side. This case is represented in figures \ref{fig:turing} (b) and \ref{fig:PD_typeI} (a). If the eigenvalues are complex conjugate at $k_\pm$, then $\Lambda_{1/2}(k_\pm)=\pm i \text{Im}(\Lambda_{1/2}(k_\pm))$ and the state curve crosses the line corresponding to $ \Trace\mathcal{G}=0$; this case is shown in figures \ref{fig:turing} (c)-(d) and \ref{fig:PD_typeI} (b) \cite{PhysRevE.103.042602, PhysRevLett.131.148301}.
As shown in figures \ref{fig:turing} (e)-(f) and \ref{fig:PD_typeI} (c), a scenario which is a mixture of the two above can also arise: for example, at small wave numbers the system shows stable oscillations, while pattern formation appears at finite values of $k>0$.
In general, the system parameters can be tuned in such a way that the local negative maximum Re$(\Lambda_2(k_c))$ becomes positive and global, thus leading to instability. This type of instability is a conserved version of the well-known  Turing or type I instability. Contrarily to the standard Turing instability, in our model, it can be attained also for $D^{\rm eff}_1=D^{\rm eff}_2$. Similarly to the type II instability, it can be stationary or oscillatory.

\begin{figure}
\centering
\includegraphics[width = 0.8\linewidth]{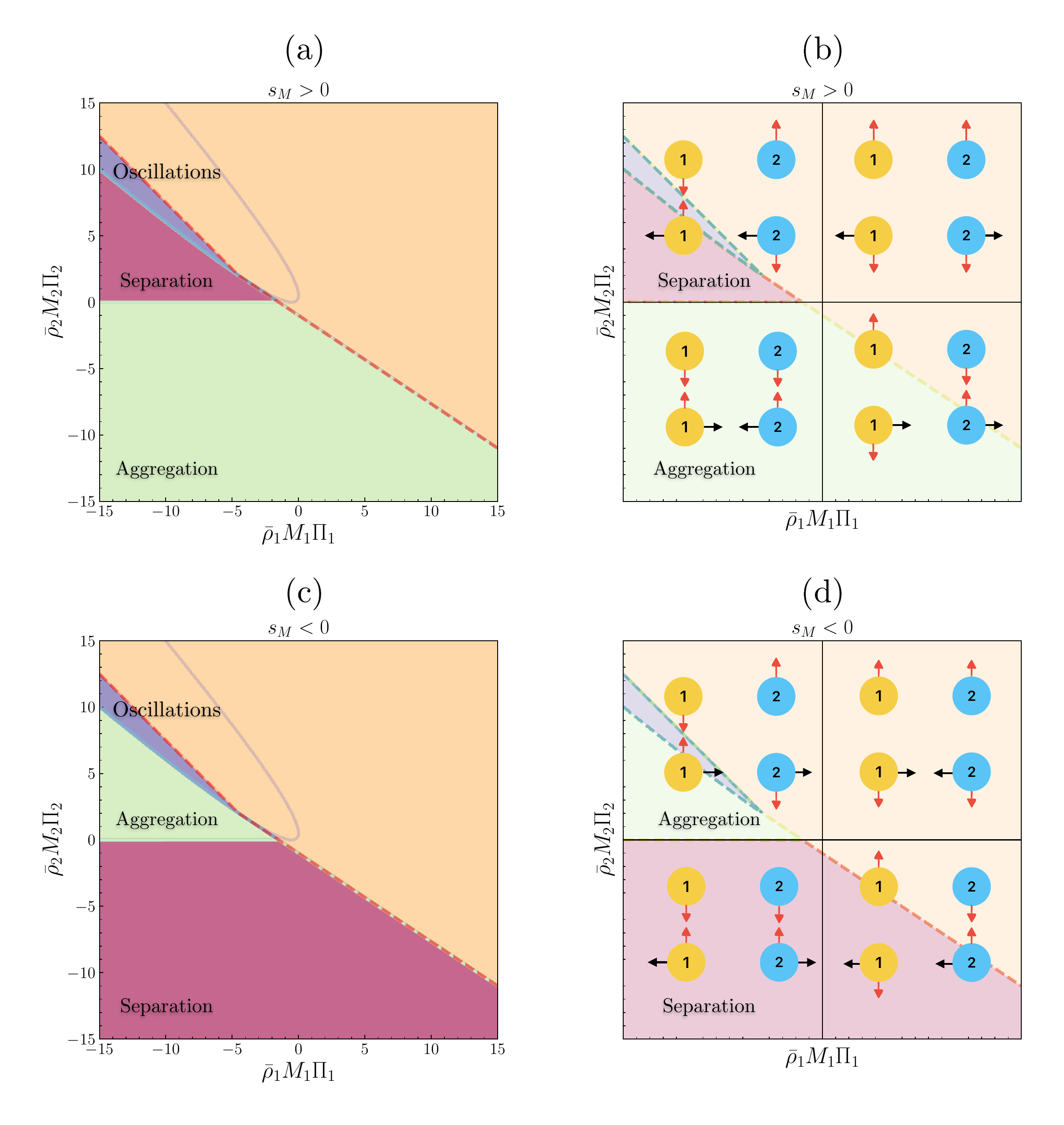}
\caption{In (a) and (c) we display the phase diagram of the effective interactions: the orange area represents the stable homogeneous phase, which is divided by the unstable phase by the red dashed line. The nature of the unstable non-oscillatory regime is captured by the relative sign $s_M$ of the effective mobilities $\bar{\rho}_a M_a$: the magenta region corresponds to the separation of the two species ($s_\rho(k)<0$), the green one to aggregation ($s_\rho(k)>0$). In (b) and (d), we show the emerging effective interactions corresponding to (a) and (c) respectively. Specifically, in (b) the effective interaction between the two type particles for $s_M>0$ and for $s_M<0$ in (d). The red arrows indicate the direction of the reciprocal interaction between particles of the same species, on the other hand, the black ones refer to nonreciprocal interactions between particles of different species. }\label{fig:effectiveinteraction}
\end{figure}

\subsection{Effective interaction}\label{sec:effint}
To grasp the mechanisms that may lead to these types of instabilities, in the next section we describe the type of nonreciprocal interactions that may arise between the two species of Janus colloids. We are now interested in looking at which of the two system modes, in the unstable regime, dominates the linear instability and how it affects the growth of the relative concentration of the two species $\delta\rho_1(\bm{k},t)/\delta\rho_2(\bm{k},t)$. In particular, its sign $s_\rho(\bm{k},t)=\sign\left(\delta\rho_1(\bm{k},t)/\delta\rho_2(\bm{k},t)\right)$ gives information on the type of instability: if negative, there will be a local depletion of one species in favor of the other one, meaning separation, otherwise local growth of concentration leads to aggregation. 

We show in appendix \ref{app:EffectiveInteraction} that $s_\rho(\bm{k},t)$ can be factorized as $s_\rho(\bm{k},t)=s_M\,s_m(\bm{k},t)$, 
and it is given by 

\begin{equation}
    s_M={\rm sign}(M_1/ M_2),\;\;\;\text{and}\;\;\;s_m={\rm sign}\left(\frac{D_1^{\rm eff}+\bar{\rho}_1 M_1 \Pi_1-D_2^{\rm eff}-\bar{\rho}_2 M_2 \Pi_2-2\sqrt{\Delta}}{M_1\Pi_1}\right).
\end{equation}
It can be easily checked that in the unstable regime of the phase diagram, for $\bar{\rho}_2 M_2\Pi_2(k)>0$ the factor $s_m(k)<0$ and one has that the sign of the instability is opposite that of the ratio of the effective mobilities  $s_\rho=-s_M$, while for $\bar{\rho}_2 M_2\Pi_2(k)<0$ the type of instability is reversed, i.e., $s_\rho=s_M$. Therefore, aggregation of the two-particle species, or equivalently $s_\rho>0$, is expected for $\bar{\rho}_2 M_2\Pi_2(k)<0$ whenever $s_M>0$ or for $\bar{\rho}_2 M_2\Pi_2(k)>0$ if $s_M<0$ (green area in figure \ref{fig:effectiveinteraction}); particles separate otherwise (magenta area in figure \ref{fig:effectiveinteraction}).
The type of effective interaction between the two species depends on the sign of the mobilities $\bar{\rho}_a M_a$ and the production rate $\Pi_a(k)$. In particular, if $\bar{\rho}_a M_a>0$ the particle of species $a$ will move towards regions where the density of the substrate $c(\bm{r},t)$ decreases, which means that the particle is attracted by consumers of the substrate with production rate $\Pi_b(k)<0$ and repelled by producers with $\Pi_b(k)>0$. The details of all the possible interactions between the two species are reported in figure \ref{fig:effectiveinteraction}. Because of the assumption $D_1^{\rm eff}>D_2^{\rm eff}$, the dynamics of the first species are faster than that of the second species, and it responds faster to the presence of chemical gradients, leading to the prevailing of effective interaction felt by the first species with respect to the second one. Referring to figure \ref{fig:effectiveinteraction}, this explains why independently of the type of effective interaction experienced by the second species, if the first species is attracted ( respectively repelled) by the second one the system displays aggregation (respectively separation).
In the unstable oscillatory regime (blue area in figure \ref{fig:effectiveinteraction}) there is an alternation of depletion and aggregation in time. 

   
The evolution of these two density perturbations can be characterized by looking at their phase and amplitude as $\delta\rho_{1,2}(\bm{k},t)e^{-k^2\Trace{\mathcal{G}}/2}=A_{1,2}\cos(\sqrt{|\Delta|}k^2t+\varphi_{1,2})$.
Albeit the two perturbations are destined to grow exponentially, in this linear approximation the relative amplitude of the oscillations $A
_1/A_2$ is constant in time, and it depends on the initial value of the perturbation.
On the other hand, the phase difference $\Delta\varphi = \varphi_1-\varphi_2$ is independent of the initial perturbation, and it is given by
\begin{equation}\label{eq:phasedifference}
\tan\Delta\varphi = -\frac{2\sqrt{|\Delta|}}{\mathcal{G}_{11}-\mathcal{G}_{22}}=-\sqrt{-1-\frac{4\bar{\rho}_1M_1\Pi_1\bar{\rho}_2M_2\Pi_2}{(\delta D +\bar{\rho}_1M_1\Pi_1-\bar{\rho}_2M_2\Pi_2)^2}}.  
\end{equation}

\begin{figure}
\centering
\includegraphics[width = \linewidth]{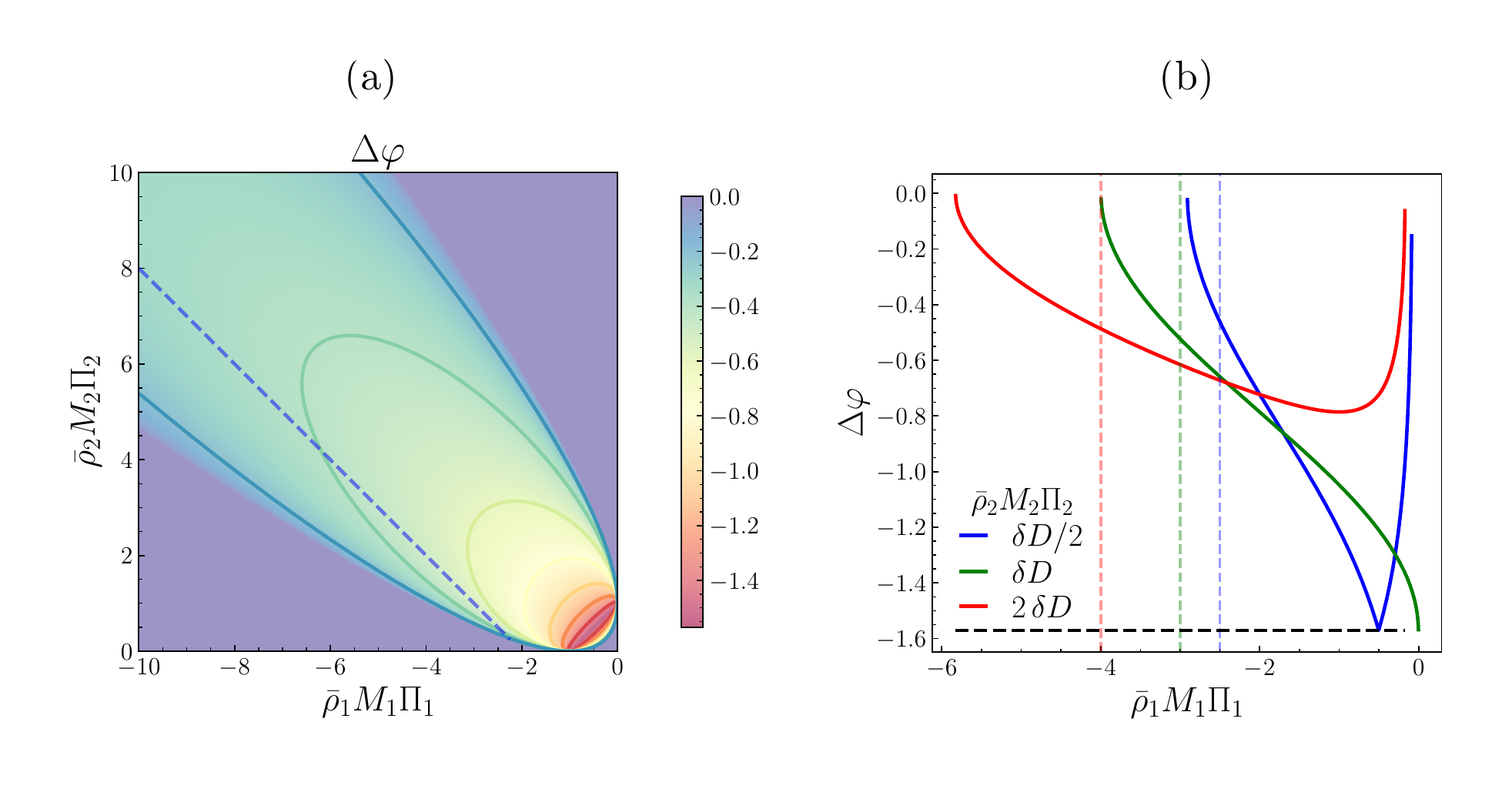}
\caption{We showcase the phase difference in equation \eqref{eq:phasedifference} between the two species in the oscillating state. (a) displays the phase difference in the region of complex eigenvalues: on the left of the dashed line, the eigenvalues are unstable, stable otherwise. (b) represents $\Delta\varphi$ for different values of $\bar{\rho}_2M_2\Pi_2$ as a function of $\bar{\rho}_1M_1\Pi_1$. The dashed vertical line separates values of $\Delta\varphi$ corresponding to unstable eigenvalues (on the left) and to stable ones (on the right). }\label{fig:phasedifference}
\end{figure}

 The phase difference, as shown in figure \ref{fig:phasedifference}, takes value in $(-\pi/2, 0)$: in the unstable region ($\Trace\mathcal{G}>0$) the phase difference is always above $-\pi/2$, while it is in quadrature only in the stable phase for $\Trace\mathcal{G}=0$, compatibly with $\bar{\rho}_1M_1\Pi_1\in(-\delta D, 0)$ and $\bar{\rho}_2M_2\Pi_2\in(0, \delta D)$.  

\section{Conclusions}
\begin{figure}
\centering
\includegraphics[width = \linewidth]{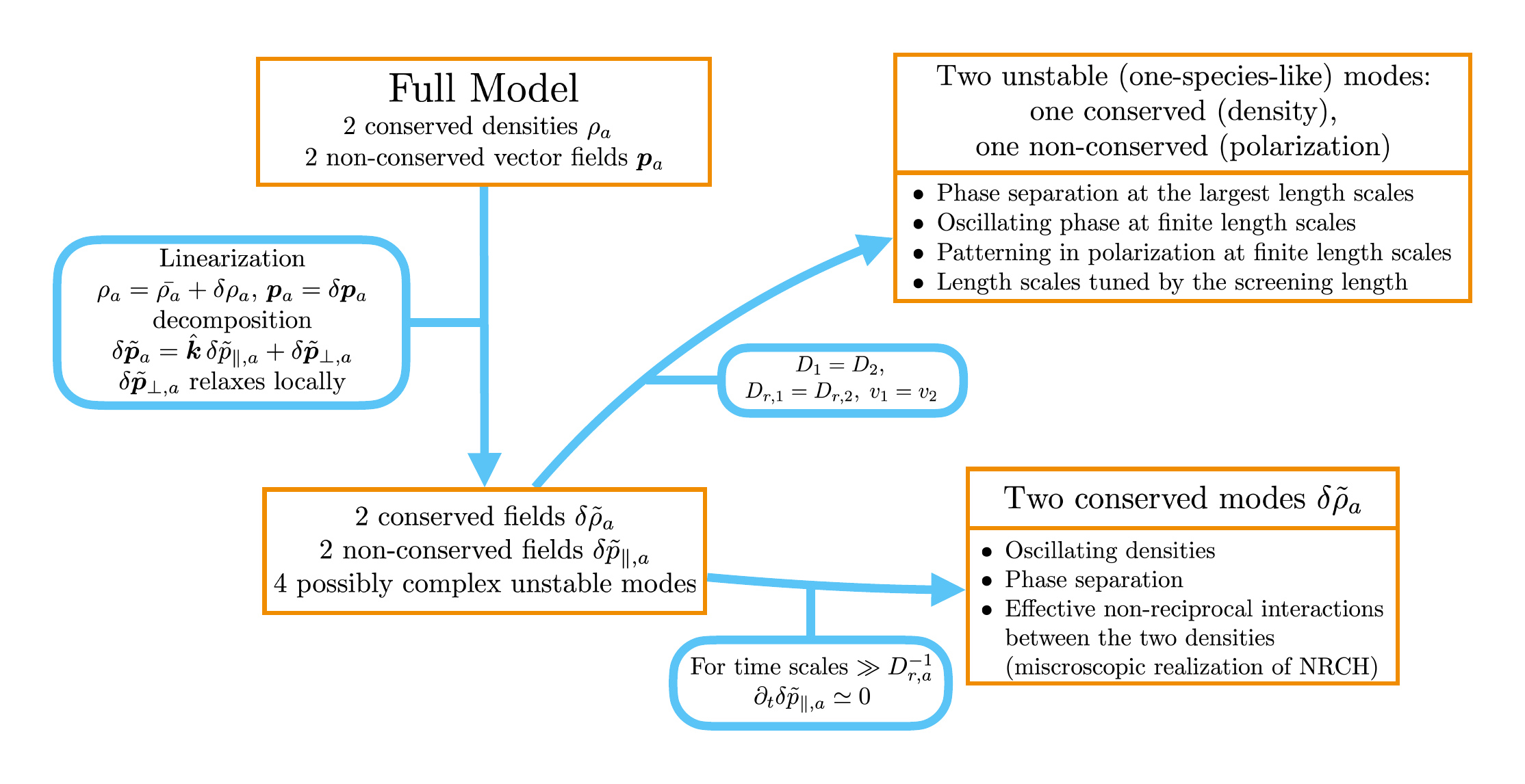}
\caption{Synoptic diagram of the main results of the paper. We schematize our work's logical flow and results. We have started from two sets of coupled nonlinear differential equations, each describing the time evolution of the density and polarization field for a given species of Janus colloids. Then, we have discussed their linearization around a homogeneous disordered state. We have split the analysis into two different parameters' regimes: i) we have first considered equal diffusivities and self-propelling velocities for the two species, ii) then the limit of fast relaxation of the polarization field.
}\label{fig:finalsketch}
\end{figure}

In this work, we have shown how effective nonreciprocal interactions arise in a collection of two types of Janus colloids coupled to the same chemical substrate. Including two species represents the minimal requirement for nonreciprocal couplings between number density fields. We refer to figure \ref{fig:finalsketch} for a synoptic sketch of our main results. First, we have introduced the corresponding single-particle dynamics, describing a set of Janus colloids that can move and re-orient along the gradient of a chemical substrate, which is produced or consumed by the colloids themselves. From this microscopic description, we have derived the hydrodynamic equations for the relevant slow modes, i.e., the particle density and the polarization field for each species, which capture the collective behavior of the system. We have derived the corresponding equations for small deviations from the spatially homogenous and orientationally disordered state, allowing us to establish the linear stability of this phase. Janus particles with chemical field-mediated effective interactions are analogous to screened Coulombic systems. The eigenvalues determining the linear stability of the system assume simple forms when the translational and rotational diffusion coefficients and self-propelling velocities are equal. Eigenvalue analysis predicts phase separation as the most robust behavior at the largest lengthscales with a diffusion coefficient whose sign is controlled both by the chemotactic drift and the angular rotation. At the scale of the screening length, phoretic coupling between number and orientation fields leads to oscillations. Oscillations appear either through the mechanism of Hopf bifurcation or when the system crosses an exceptional point. Two pairs of complex eigenvalues appear in the most generic case, a scenario where two density fields and two orientation fields undergo oscillations.

For large rotational diffusivity, the polar fields simply follow the density gradients. In this scenario, the polarization degrees of freedom can be expressed in terms of the density by an adiabatic approximation. The resulting equations for the two density fields are linearized around the homogeneous solution yield expressions for the interaction coefficients featuring wave-vector-dependent activity and mobility coefficients. The interplay among the associated effects leads to a nonreciprocal interaction between particles of the same or different species leading to aggregation or separation phenomenon. We have found different types of short-wavelength, stationary, or oscillatory instabilities~\cite{Thiele2023,BenoitYu2023}. In particular, contrary to what happens in standard Turing instability, we get such instability even in the case of equal effective diffusivities, as a consequence of the wave-vector-dependence of the phoretic parameters. The analysis provides a direct link to the nonreciprocal Cahn-Hilliard model (NRCH)~\cite{NRCH, you2020nonreciprocity}, and suggests that nonreciprocity should be incorporated in the surface tension to approach a more complete theoretical framework for scalar active densities. Recent papers have elucidated that NRCH serves as a minimal model for known systems such as active-passive mixtures, mass-conserving reaction-diffusion systems, and active gels~\cite{brauns2023nonreciprocal,Thiele2023}. To the best of our knowledge, our paper is the first work that starts from the microscopic model of a chemically active swimmer including self-propulsion and orientational dynamics to enumerate the various contributions to effective intra-species and inter-species interactions (both reciprocal and nonreciprocal) in terms of single-particle phoretic or enzymatic activity~\cite{Agudo-Canalejo2018_1,Agudo-Canalejo2018_2} and mobilities thus providing several routes to realizing the NRCH. In general, the speed of self-propulsion and chemical activity could depend on the number densities thus providing a route observing the effect of nonlinearity in nonreciprocal interactions~\cite{saha2022effervescent}. 

An explicit manifestation of non-reciprocity is to enforce reactive couplings between thermodynamic fluxes that should not be so coupled. Gradients of chemical potentials should be coupled dissipatively with symmetric coefficients~\cite{DeGrootMazur}, while velocity fields and density are advectively coupled through coefficients of the same magnitude~\cite{Halperin_Hohenberg,PrawarDadhichi2018,Das2002,das2004nonequilibrium}. Our system presents two scenarios when Onsager's principle is violated in both forms -- cross-couplings between densities of different species and coupling between the longitudinal component of the polarization with density. 

Taking cues from the analysis, it is important to explore the full dynamical behavior of the system through a solution of the equations presented here or in agent-based simulations of the microscopic model. A condensation of the longitudinal component of the polarity only, a state called asters in \cite{PhysRevE.89.062316} deserves a thorough study examining questions such as long-range correlations \cite{PhysRevE.58.4828, PhysRevLett.75.4326, TONER2005170}. We expect a proliferation of defects for a single species, somewhat similar to and yet distinct (i.e. occurring through a different mechanism) from the defects observed in a Malthusian flock~\cite{BesseChateSolon}. For two species, where the stability analysis shows all modes to be unstable, we speculate a state with interacting defects. Our work can be generalized in several ways -- to multicomponent mixtures interacting with several substrates~\cite{VincentPRL}, coupling the mixture to a momentum-conserving fluid~\cite{LushiPhysRevE.98.052411, VinzeMichelinPhysRevFluids.9.014202}, dynamics at an interfaces and close to boundaries~\cite{Malgaretti2016,Uspal2015,Popescu2017}, and entropy production~\cite{bebon2024thermodynamics}. Several aspects of our work can be generalized to other versions of tactic systems - whether it is phototaxis~\cite{LozanoTurning2016} or quorum sensing~\cite{TailleurTactic_PRL}. Finally, our work illustrates that the physics of active mixtures represents a rich area of research and presents many predictions that can be tested in experiments~\cite{Wang2023,SmallSharan2023,Sánchez2015}.

\newpage
\appendix

\section{Coarse-graining of the Langevin dynamics}\label{app:CG}

In this section, we show how to derive equations \eqref{eq:rho} and $\eqref{eq:ptruncated}$. We start by considering the evolution of $\mathcal{P}_a(\bm{r},\bm{n},t)$, the probability density of a particle of species $a\in\{1,\dots,n\}$ to be at position $\bm{r}$ and orientation $\bm{n}$ at time $t$, defined as
\begin{equation}\label{eq:Padef}
\mathcal{P}_a(\bm{r},\bm{n},t)=\Bigg\langle\sum_{i=1}^{N_a}\delta(\bm{r}-\bm{r}_{a,i})\delta(\bm{n}-\bm{n}_{a,i})\Bigg\rangle,
\end{equation}
for a set of $N_a$ Janus colloids of species $a$, where the average is taken with respect to the noise contribution to the dynamics.
Then, considering that the process in equation \eqref{eq:langevin} follows Stratonovich convention of stochastic calculus, it can be easily shown that $\mathcal{P}_a(\bm{r},\bm{n},t)$ satisfies the following Fokker-Planck equation
\begin{equation}\label{eq:FP}
\begin{aligned}
\partial_t \mathcal{P}_a(\bm{r},\bm{n},t)=&-\nabla\cdot\left\{\left[v_a\bm{n}-\mu_a\nabla c\right]\mathcal{P}_a(\bm{r},\bm{n},t)\right\}+D_a\nabla^2\mathcal{P}_a(\bm{r},\bm{n},t)\\
&-\mathcal{R}\cdot\left[\Omega_a\left(\bm{n}\times \nabla c\right)\mathcal{P}_a(\bm{r},\bm{n},t)\right]+D_{r,a}\mathcal{R}^2\,\mathcal{P}_a(\bm{r},\bm{n},t),
\end{aligned}
\end{equation}
where $\mathcal{R}\equiv \bm{n}\times\nabla_n$ is the orientational gradient operator.
The first line on the right-hand side of equation \eqref{eq:FP} describes the contribution to the probability flux due to drift and diffusion of the particle position, whereas the second line to alignment interaction and diffusion of its orientation.

In order to find an equation for $\rho_a$ and $\bm{p}_a$, we restrict our analysis to the first and second moments of the orientation $\bm{n}$ by closing the corresponding hierarchy of infinite many equations for the moments generated from equation \eqref{eq:FP}.
We start by integrating equation \eqref{eq:FP} with respect to $\bm{n}$, which leads to the time evolution of density of the particles $\rho_a(\bm{r},t)\equiv\int_{|\bm{n}|=1}\mathrm{d}\bm{n}\,\mathcal{P}_a(\bm{r},\bm{n},t)$ of species $a$ in equation \eqref{eq:rho}. 
Similarly, one can calculate the dynamics of the polarization field $\bm{p}_a(\bm{r},t)\equiv\int_{|\bm{n} |=1}\mathrm{d}\bm n\,\bm{n}\,\mathcal{P}_a(\bm{r},\bm{n},t)$ as
\begin{equation}\label{eq:p}
\partial_t\bm{p}_a=-\nabla\cdot\left[v_a \mathbb{Q}_a-\mu_a\bm{p}_a\nabla c\right]-\frac{v_a}{3}\nabla\rho_a+\Omega_a\left[\frac{2}{3}\rho_a\nabla c-\nabla c\cdot \mathbb{Q}_a\right]+(D_a\nabla^2-2D_{r,a})\bm{p}_a,
\end{equation}
where we have introduced the nematic tensor $\mathbb{Q}_a(\bm{r},t)=\int_{|\bm{n}|=1}\mathrm{d}\bm{n}\,\left(\bm{n}\bm{n}-\mathbb{I}/3\right)\mathcal{P}_a(\bm{r},\bm{n},t)$.

In the evaluation of equation \eqref{eq:p}, which is obtained by multiplying by $\bm{n}$ equation \eqref{eq:FP} we have calculated the following non trivial integrals:

\begin{itemize}
    \item[i)] The first contribution that we consider is the $i$-component of  

\begin{equation}
\begin{aligned}
    -\int_{|\bm{n}|=1}\mathrm{d}\bm{n}\, \, n_i\,\mathcal{R}\cdot\left[(\bm{n}\times\nabla c) \,\mathcal{P}_a(\bm{r},\bm{n},t)\right]&=\int_{|\bm{n}|=1}\mathrm{d}\bm{n}\, \, (\mathcal{R}_j n_i)\,(\bm{n}\times\nabla c)_j \,\mathcal{P}_a(\bm{r},\bm{n},t)\\
    &=-(\partial_m c)\int_{|\bm{n}|=1}\mathrm{d}\bm{n}\, \,  n_k n_l(\delta_{il}\delta_{km}-\delta_{im}\delta_{kl}) \,\mathcal{P}_a(\bm{r},\bm{n},t)\\
    &=-(\partial_j c)\,\int_{|\bm{n}|=1}\mathrm{d}\bm{n}\, (n_i n_j-\delta_{i j})\mathcal{P}_a(\bm{r},\bm{n},t)\\
    &=\left[\frac{2}{3}\rho_a(\bm{r},t)\nabla c-\nabla c\cdot\mathbb{Q}_a(\bm{r},t)\right]_i.
\end{aligned}
\end{equation}
On the right hand side of the first line  we exploit the fact that $\mathcal{R}_i$ satisfies the typical properties of gradient operator (that is, chain differentiation rule and hence integration by part), and the relation $\mathcal{R}_in_j=-\epsilon_{ijk}n_k$. The second line is obtained by contraction of the Levi Civita symbol $\epsilon_{jik}\epsilon_{jlm}=\delta_{il}\delta_{km}-\delta_{im}\delta_{kl}$. The last equality follows from the definition of the particle density $\rho_a$ and the 3-dimensional nematic tensor $\mathbb{Q}_a$.

\item[ii)] The second non-trivial term contributing to the $\bm{p}_a$ dynamics is given by the one associated with angular diffusion, given by 

\begin{equation}
\begin{aligned}
    \int_{|\bm{n}|=1}\mathrm{d}\bm{n}\,\,n_i \,\mathcal{R}^2\,\mathcal{P}_a(\bm{r},\bm{n},t)&=\int_{|\bm{n}|=1}\mathrm{d}\bm{n}\,\,\mathcal{P}_a(\bm{r},\bm{n},t)\,\mathcal{R}^2 n_i=-2p_{a,i},\\
\end{aligned}
\end{equation}
that simply follows from the relation  $\mathcal{R}^2n_i=-2n_i$.
\end{itemize}
Closure in the moment expansion can be attained by
considering the case where the nematic order is negligible $\mathbb{Q}_a=0$.
This truncation of the hierarchy of $\bm{n}$-moments in \eqref{eq:FP} simplifies equation \eqref{eq:p} for $\bm{p}$ to \eqref{eq:ptruncated}.

\subsection{ Microscopic origin of phoretic interactions}

The phoretic couplings $\mu_a$, $\Omega_a$, and the production rates $\alpha_a$, and $\beta_a$ can be expressed in terms of microscopic parameters describing the geometric distribution of the mobility $\mu^{(a)}$ and activity $\alpha^{(a)}$ on the surface of spherical Janus particles.
If we restrict to a mobility $\mu^{(a)}$ and activity $\alpha^{(a)}$ that is axis-symmetric with respect to $\bm{n}_a$, they are parametrized along the surface of the Janus particle only via $\cos\theta$, where $\theta$ denotes the angle with respect to the symmetry axis. It is then convenient to expand $\mu^{(a)}(\cos\theta)$ and $\alpha^{(a)}(\cos\theta)$ in Legendre polynomials according to

\begin{equation}\label{eq:Legendrexpansion}
    \mu^{(a)}(\cos\theta)=\sum_{m=0}^\infty P_m(\cos\theta)\mu^{(a)}_m,\quad\quad\alpha^{(a)}(\cos\theta)=\sum_{m=0}^\infty P_m(\cos\theta)\alpha^{(a)}_m,
\end{equation}
where $P_m(\cos\theta)$ is the $m-$th degree Legendre polynomials, and the coefficients $\mu^{(a)}_m$ and $\alpha^{(a)}_m$ are given by 
\begin{equation}
    \mu^{(a)}_m=\left(m+\frac{1}{2}\right)\int_{0}^{\pi}\mathrm{d}\theta\,\sin\theta\,P_m(\cos\theta)\mu^{(a)}(\cos\theta),\quad \quad \alpha^{(a)}_m=\left(m+\frac{1}{2}\right)\int_{0}^{\pi}\mathrm{d}\theta\,\sin\theta\,P_m(\cos\theta)\alpha^{(a)}(\cos\theta).
\end{equation}

It can be shown that the velocity $\bm{v}_{a}$ and angular velocity $\bm\omega_a$ \cite{golestanian2007designing, kanso2019phoretic} due to phoretic interaction with the chemical substrate are given by

\begin{equation}
    \bm{v}_a(\bm{r},t)=-\mu^{(a)}_0\nabla c(\bm{r},t)+\frac{3}{10}\mu^{(a)}_2\left(\bm{n}_a\bm{n}_a-\frac{\mathbb{I}}{3}\right)\cdot \nabla c(\bm{r},t),\quad\quad \bm\omega_a = -\frac{3\mu^{(a)}_1}{4 R_a}\bm{n}_a\times\nabla c(\bm{r},t),
\end{equation}
where $R_a$ is the radius of the colloid. From last equation we immediately read the coefficients $\mu_a = \mu^{(a)}_0$ and $\Omega_a = -3\mu^{(a)}_1/(4R_a)$ appearing in equation \eqref{eq:langevin}, while $\mu^{(a)}_2=0$ if we assume hemispherically coated Janus colloids.

\subsection{Equation for the chemical substrate}
The set of dynamical equations is completed by the evolution of the substrate density field  $c(\bm{r},t)$, which is given by
\begin{equation}\label{eq:capp}
\partial_t c-D_c(\nabla^2 -\kappa^2)c=\Pi(\bm{r},t)=\sum_{a=1}^n\left[\alpha_a\rho_a-\beta_a\nabla\cdot\bm{p}_a+O(R_a^5)\right],
\end{equation}
where $R_a$ is the radius of the $a-$species Janus particle, and $\Pi$ the local production rate of chemical substrate.

The average production rate $\Pi(\bm{r},t)$ can be expressed by integrating the contribution coming from the local production rate  $\alpha^{(a)}(\cos\theta_{i_a})$, parametrized by $\theta_{i_a}$, the angle between a point on the surface of each $i_a\in\{1,\dots,N_a\}$ Janus colloid and its axis, i.e.,
\begin{equation}\label{eq:Piexpansion}
\begin{aligned}
    \Pi(\bm{r},t) &= \Bigg\langle\sum_{a=1}^n\sum_{i_a=1}^{N_a}\int_{|\bm{R}_{i_a}|=R_a}\delta(\bm{r}-\bm{r}_{i_a}-\bm{R}_{i_a})\alpha^{(a)}(\cos\theta_{i_a})\Bigg\rangle\\
    &=\Bigg\langle\sum_{a=1}^n R_a^2 \sum_{i_a=1}^{N_a}\sum_{m=0}^{\infty}\alpha^{(a)}_m\int_{0}^{\pi}\mathrm{d}\theta_{i_a}\,\sin\theta_{i_a}\int_{0}^{2\pi}\mathrm{d}\varphi_{i_a}\,P_m(\cos\theta_{i_a})\left[1-R_a\hat{\bm{r}}_{i_a}\cdot\nabla+O(R_a^2)\right]\delta(\bm{r}-\bm{r}_{i_a})\Bigg\rangle\\
    &=\sum_{a=1}^n 4\pi R_a^2\Bigg\langle \sum_{i_a=1}^{N_a}\sum_{m=0}^{\infty}\alpha^{(a)}_m\left\{\delta_{m,0}-\frac{R_a}{3}\delta_{m,1}(\bm{n}_{i_a}\cdot\nabla)+O(R_a^2)\right\}\delta(\bm{r}-\bm{r}_{i_a})\Bigg\rangle\\
    &=\sum_{a=1}^{n}4\pi R_a^2\left[\alpha^{(a)}_0\,\rho_a(\bm{r},t)-R_a\frac{\alpha^{(a)}_1}{3}\,\nabla\cdot\bm{p}_a(\bm{r},t)+O(R_a^2)\right],
\end{aligned}
\end{equation}
where $\bm{r}$ is a point on the surface of the particle, $\bm{r}_{i_a}$ the location of the center of the $i_a$-th particle of species $a$, and $\bm{R}_{i_a}=\bm{r}-\bm{r}_{i_a}$ is a point on the particle in the particle's reference frame. In the second line, we expand the activity $\alpha^{(a)}$ in Legendre polynomials according to equation \eqref{eq:Legendrexpansion}, we Taylor-expand the Dirac delta for $R_a\ll r$, and we make the surface integral explicit. The last two lines follow from integration and the definition of $\rho_a$ and $\bm{p}_a$. From the above equation, we read 

\begin{equation}
\alpha_a=4\pi R_a^2 \alpha^{(a)}_0, \quad\quad \beta_a = \frac{4}{3}\pi R_a^3 \alpha^{(a)}_1.    
\end{equation}

\section{Construction of the phase diagram}\label{app:conpd}
We now discuss the steps that we follow to construct the stability phase diagram presented in the main text. We consider a two-dimensional linear with the associated two eigenvalues
\begin{equation}
\begin{aligned}
    \Lambda_{1,2}=\frac{\Trace \mathcal{G}}{2}\mp \sqrt{\Delta},\;\;\;\text{with}\;\;\;\Delta=\left(\frac{\Trace \mathcal{G}}{2}\right)^2-\Det \mathcal{G}.
\end{aligned}
\end{equation}
We now summarise the results of the eigenvalue analysis in a stability diagram. Generically, we can distinguish three significant regions denoted by $R_{\Delta,1,2}$ delimited by curves $C_{\Delta,1,2}$:
\begin{equation}\label{eq:regions}
\begin{split}
& R_\Delta: \Delta\le 0, \\
& R_1 :\Trace\mathcal{G}(k)=\Lambda_1+\Lambda_2<0,\\
& R_2: \Det\mathcal{G}(k)=\Lambda_1\Lambda_2>0, 
\end{split} \quad\quad
\begin{split}
& C_\Delta : \Delta = 0, \\
& C_1 : \Trace\mathcal{G}=0,\\
& C_2:\Det\mathcal{G}=0.
\end{split}
\end{equation}
$R_{\Delta,1,2}$ identify the regions with real eigenvalue, positive trace of $\mathcal{G}$ and positive determinant $\rm{det} \mathcal{G}$ respectively. The intersection of $R_1$ and $R_2$ identifies the stability region of the system. The region $R_\Delta$ corresponds to the region of complex eigenvalues. The boundary $C_\Delta$ is the line of \textit{exceptional points}, where the eigenvalues are equal. 
Note that $R_\Delta$ is always contained in $R_2$, implying that by crossing the portion of $C_1$ belonging to $R_\Delta$ the stability of the system changes. 
To illustrate one application of the considerations above, as an example, we construct in detail the phase diagram shown in figure \ref{fig:phasediagram_dim}. 

We can easily determine the topology of these regions in the plane of the parameters $(\bar{\rho}_1 M_1\Pi_1, \bar{\rho}_2 M_2\Pi_2)$ while keeping $D_2^{\rm eff}$ and $\delta D>0$ fixed. 
It can be simply checked from equation \eqref{eq:eigenvalues} for $\Delta$, that this curve is a parabola defined for $\bar{\rho}_1 M_1\Pi_1\le 0$ with symmetry axis $\bar{\rho}_2 M_2\Pi_2+\bar{\rho}_1 M_1\Pi_1=0$ and vertex $(-\delta D/4,\delta D/4)$. It is represented in figure \ref{fig:phasediagram_dim} by the blue dashed curve: the interior of the parabola corresponds to complex eigenvalues, while its complementary $R_\Delta$ to the region of real eigenvalues. 

Along the curve $C_1$ defined by the line $ \bar{\rho}_2 M_2\Pi_2+\bar{\rho}_1 M_1\Pi_1+D_1^{\rm eff}+D_2^{\rm eff}=0$, the eigenvalues $\Lambda_{1,2}=\pm\sqrt{\bar{\rho}_1 M_1\Pi_1\bar{\rho}_2 M_2\Pi_2}$ are equal and opposite. A complex conjugate pair of $\Lambda_{1,2}$ change their sign on crossing $C_1$. Thus $C_1$ lying in $R_{\Delta}$ represents points where Hopf bifurcation occurs dividing it into two regions - one where oscillations grow and the other where they decay. A pair of real $\Lambda_1$ are equal and opposite on the curve $C_1$, which means it lies in the region where one of the eigenmodes is unstable. Moreover, $C_1$ is parallel to the axis of symmetry of $C_\Delta$, such that they intersect only at one point $P_\star=\left(-(D_1^{\rm eff})^2/\delta D,(D_2^{\rm eff})^2/\delta D\right)$ where the eigenvalues are both null (blue star in figure \ref{fig:phasediagram_dim}). In figure \ref{fig:phasediagram_dim} this former branch of $C_1$ is represented by the pale green semi-line that originates from $P_\star$.  

The expression of the curve $C_2$ is given by line $ \bar{\rho}_1 M_1\Pi_1/D_1^{\rm eff}+\bar{\rho}_2 M_2\Pi_2/D_2^{\rm eff}+1=0$ and its corresponding eigenvalues read $\Lambda_1 = \Trace\mathcal{G}$ and $\Lambda_2 = 0$.
Note that $C_2$ is tangent to $C_\Delta$ at $P_\star$ where also  $\Lambda_1=0$. Thus, as in the case of $C_1$, the point $P_\star$ splits $C_2$ into two semi-lines with different behavior: one for $\bar{\rho}_1 M_1\Pi_1>(D_1^{\rm eff})^2/\delta D$ where $\Lambda_1$ is stable (red semi-line in figure \ref{fig:phasediagram_dim}) and the other that lays in the (real) instability region. Indeed, for $\bar{\rho}_1 M_1\Pi_1>(D_1^{\rm eff})^2/\delta D$ above $C_2$ we have real and stable eigenvalues, that become unstable below. 

Similar considerations allow us to obtain the phase diagrams in figure \ref{fig:PDtwospecieslikeone}.

\section{ Details of single species dynamics }\label{app:OneSpecies}
Already at its linear description in equation \eqref{eq:rhoplinear}, it is a very hard task to deal with the complexity of the multi-species dynamics. For this reason, it is useful to get some insights from the single component case. Indeed, as we have anticipated in section \ref{sec:twospecies}, it allows us to qualitatively understand certain simplified regimes of the multi-species case. For the one species case, the linearized dynamics of a perturbation to the homogeneous solution $\rho(\bm{r},t)=\bar\rho$, $\bm{p}(\bm{r},t)=0$, and $c(\bm{r},t)=\bar{c}=\alpha\bar\rho/(D_c\kappa^2)$  is then given by

\begin{equation}\label{eq:1spdeceq}
\begin{aligned}
&\partial_t\drk=-k^2\left[\mu\bar{\rho}\,\dck+D\,\drk\right]-ik v\, \dwk,\\
&\partial_t \dwk=-\frac{ik}{3}\left[v\,\drk-2\Omega\bar{\rho}\, \dck\right]-(2D_r+Dk^2) \,\dwk,\\
&\dck=\frac{\alpha\drk-ik\beta \dwk}{D_c\left(\kappa^2+k^2\right)},\quad 
\partial_t \dperk=-(2D_r+Dk^2)\dperk.\\
\end{aligned}
\end{equation}
Note that, to ensure the physical requirement $\bar{c}>0$ of positive substrate density, this description entails a positive $\alpha>0$ and $\kappa\neq0$: the substrate is on average created by the Janus colloids, and its dynamics has to be screened.
Equation \eqref{eq:1spdeceq} tells us that perturbation to the homogeneous density profile $\drk$ is influenced only by the component along $\bm{k}$ of the polar field, while its orthogonal contribution $\dperk$ is exponentially suppressed over time due to angular diffusion. For this reason, we restrict our analysis to the evolution of $\delta\rho$ and $\delta p_\parallel$.

Accordingly, we can express the linearized dynamics of the system as 

\begin{equation}\label{eq:onespecieseq}
    \partial_t(\drk,\delta p_\parallel(\bm{k},t))^T=\mathcal{G}(\bm{k})(\drk,\delta p_\parallel(\bm{k},t))^T,
\end{equation}
where we identify the one-species dynamical matrix as
\begin{equation}\label{eq:Done}
    \mathcal{G}(\bm{k})=-\left(
\begin{array}{cc}
Dk^2+K(k)\frac{\bar{\rho}\mu\alpha}{D_c} & ik\left(v-K(k)\frac{\bar{\rho}\mu\beta}{D_c}\right) \\[7pt]
ik\left(\frac{v}{3}-K(k)\frac{2\bar{\rho}\Omega\alpha}{3k^2D_c}\right) & 2D_r+Dk^2-K(k)\frac{2\bar{\rho}\Omega\beta}{3D_c}\\
\end{array}\right),
\end{equation}
with $K(k)\equiv k^2/(k^2+\kappa^2)$.
We recall that, at the single particle level, the phoretic drift $\mu$ and alignment interaction $\Omega$ determine how the Janus particles respond to the gradient distribution of the substrate $\nabla c$: for positive values of $\mu$ the particles escape from high $c$-concentration regions, while they point towards high concentration region for $\mu<0$; for $\Omega>0$ particles align along $\nabla c$ (high concentration) and anti-align otherwise. For $\beta>0$ more substrate particles are produced in the orientation $\bm{n}$ direction, they are consumed in the opposite case. 
\begin{figure}
\centering
\includegraphics[width =\linewidth]{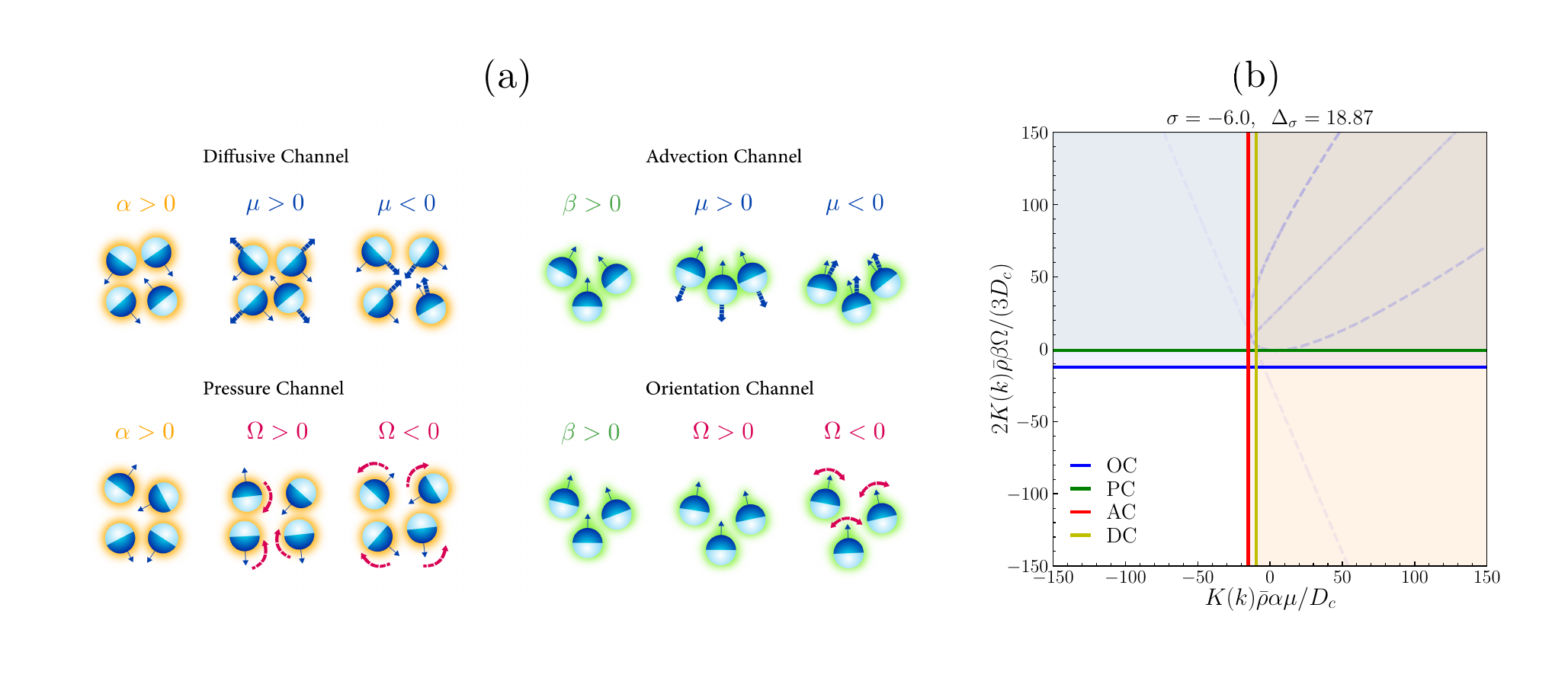}
\caption{In panel (a), we show all the possible channels of effective interactions in $\mathcal{G}$ due to phoretic effects. On the top left we represent the effect of phoresis on the effective diffusivity of the linearized dynamics corresponding to $\mathcal{G}_{11}$. The first column identifies the type of perturbation to the homogenous phase: in this case a local collection of Janus colloids with random orientation $\bm{n}$ (black arrow) producing the chemical substrate in a homogeneous way (yellow area) in a concentration higher than $\bar{\rho}$. In the second and third columns, the effect of the chemotactic drift $\mu\nabla c$ is shown, whose orientation is represented by the blue arrows. Similarly, in the top right corner, we account for the advective channel due to chemotactic drift and asymmetric production chemical described by the parameter $\beta$ (green area). On the bottom left, we display the effect of $\alpha$ and the chemotactic alignment (red arrows) to the pressure channel. Finally, the bottom right panel sketches the effect of the effective interactions on rotational diffusion. In panel (b), we show the regions of the phase diagram - similar to the one in figure \ref{fig:PDtwospecieslikeone} - where the elements of the dynamical matrix $\mathcal{G}_{ij}$ lead to a change of sign of the diffusivity and velocity positive terms at the single phoretic channel level, that is, when they are interpreted as the response coefficient of the $i$-field to the solely $j$-field perturbation. Here we have defined the auxiliary parameters $\sigma = \alpha/\beta$ and $\Delta_\sigma(k)=2D_r-v[\sigma+k^2/(3\sigma)]$, that together with the combination of parameters on the axes of the phase diagram, uniquely characterize the system.}\label{fig:channels}
\end{figure}

Therefore, we can interpret the $\mathcal{G}_{ij}$ element of the dynamical matrix $\mathcal{G}$ as the response coefficient of the $i$-field to a $j$-field small perturbation, where $i=1$ and $i=2$ identify respectively the density and polar field fluctuations. Note that the phoretic contribution to $\mathcal{G}$ appears only via one of the two production rates $\alpha$ and $\beta$, and only one of the chemotactic interactions $\mu$ and $\Omega$. This is a consequence of the linearized dynamics and it allows us to consider one type of phoretic effect at the time to the $i$-field, depending on the nature of the $j$-field perturbations to the homogeneous phase. To better understand the effect of these response coefficients we refer to figure \ref{fig:channels}, where we display the effect of these (linearized) effective interactions on the stability of the homogeneous phase:  
\begin{itemize}
    \item The first element $\mathcal{G}_{11}$ measures how small density fluctuations are amplified or suppressed while considering no perturbation to its disordered orientational component. This channel of interaction is associated with the effect of the homogeneous production of chemicals, parametrized by $\alpha$, and the chemotactic drift $\mu$. This \textit{diffusion channel} (shown on the top left of figure \ref{fig:channels}(a)) stabilizes the homogeneous phase for $\mu>0$ because of chemorepulsive interactions, whereas it can lead to aggregation in the case of effectively chemoattractive ones for $\mu<0$. In figure \ref{fig:channels} (b) we mark in yellow the region of parameters where $\mathcal{G}_{11}$ becomes negative: phoretic interaction may lead to a negative effective diffusivity.
    \item The second channel, described by $\mathcal{G}_{12}$, tells us how a small perturbation to the orientational disorder can affect the density in terms of $\beta\mu$. To fix ideas we set $\beta>0$ (top right of figure \ref{fig:channels} (a)), corresponding to the asymmetric production of chemicals along the self-propelling direction 
    $\bm{n}$. If we consider as a perturbation of the polar field the local alignment of a set of Janus colloids, we get that the effect of chemorepulsion tends to reduce the effective self-propelling velocity, whereas it increases for $\mu<0$. 
    We refer to this channel of interaction as the \textit{advective} one, since here phoretic interactions contribute to the advective part of  the density dynamics. The red region in figure \ref{fig:channels} (b) corresponds to $\mathcal{G}_{12}/(ik)$ negative, i.e. when the effective self-propelling velocity becomes negative.
    \item The third channel of interaction, associated to $\mathcal{G}_{21}$ (bottom left in figure \ref{fig:channels} (a)), can be interpreted as a \textit{pressure} term, in analogy with the case of Toner Tu type of equations \cite{TONER2005170,PhysRevE.58.4828,PhysRevLett.75.4326}. In particular, self-propulsion acts as a mechanism to restore orientational disorder due to possible local perturbation of the density field. The phoretic contribution to this mechanism is represented by $\Omega\alpha$, that is to the homogenous production of chemicals and the alignment interaction. If $\Omega>0$ Janus particles point towards high concentration regions of chemical leading to larger values of $\delta  p_\parallel$ and possibly to aggregation of particles. On the other hand, if $\Omega<0$ particles point away from high concentration regions, thus  destroying local order. Also in this case phoretic interaction may lead to a negative value of the net velocity $\mathcal{G}_{21}/(ik)$ (green area in figure \ref{fig:channels} (b)). Indeed, in the limit of fast relaxation of the polar field, e.g., $D_r+Dk^2\gg 2K(k)\Omega\beta/(3D_c)$ ($\mathcal{G}_{22}<0$), we get $\dwk \simeq -(\mathcal{G}_{21}/\mathcal{G}_{22})\drk$ and $\partial_t\drk = (\Det{\mathcal{G}}/\mathcal{G}_{22})\drk$. If (for simplicity) we set $\beta=0$, it is then apparent that the stability of the effective diffusivity $\Det{\mathcal{G}}/\mathcal{G}_{22}=\mathcal{G}_{11}-v[vk^2/3-K(k)2\Omega\alpha/(3D_c)]/(2D_r+Dk^2)$ depends crucially on the sign and intensity of $\mathcal{G}_{21}$.
    \item Finally, the term $\mathcal{G}_{22}$ identifies the contribution of alignment interaction to the effective \textit{orientational} diffusivity. Namely, it carries information about how the polar field is affected by local orientational order. As for $\mathcal{G}_{12}$ we first set $\beta>0$ (bottom right of figure \ref{fig:channels} (a)): in the case of locally ordered particles, for $\Omega>0$ the particles are most likely be already aligned to high concentration regions thus creating aggregation, alignment, and instability in the  homogeneous phase; for $\Omega<0$ the particles start to rotate since chemical is produced in correspondence of $\bm{n}$ while the same $\bm{n}$ tries to escape high concentration regions, thus stabilizing disordered configurations. In general, as $\beta\Omega>0$ phoretic interactions favor local orientational order, while they stabilize the homogeneous phase for $\beta\Omega<0$. The blue region in figure \ref{fig:channels} (b) denotes the region of parameters where the effective orientational diffusivity becomes negative because of phoretic interactions.
\end{itemize}

It is worth noting that all the phoretic contributions in $\mathcal{G}$ are proportional to $K(k)\bar{\rho}/D_c$: their contributions become stronger for a higher concentration of particles and increasing values of substrate screening length $1/\kappa$.
The region of parameters corresponding to the instability of each of these channels is highlighted in figure \ref{fig:channels} (b). Then, the full phase diagram (the light blue shaded curves in figure \ref{fig:channels} (b)) is not just given by the superposition of the four individual response coefficients $\mathcal{G}_{ij}$, but from their interplay. This mixture of different channels is captured by the eigenvalues of the dynamical matrix $\mathcal{G}(k)$. They can be expressed via the trace $\Trace{\mathcal{G}}$ and determinant  $\Det{\mathcal{G}}$ of $\mathcal{G}$ as

\begin{equation}\label{eq:eigone}
\begin{aligned}
    \Lambda_{1/2}(k)&=\frac{\Trace\mathcal{G}(k)\mp\sqrt{\left(\Trace\mathcal{G}(k)\right)^2-4\Det\mathcal{G}(k)}}{2}\\
    &=-\left[D_r+k^2D+\frac{K(k)}{2}\frac{\bar\rho}{D_c}\left(\mu\alpha-\frac{2}{3}\Omega\beta\right)\pm\sqrt{\Delta(k)}\right],\\
\end{aligned}
\end{equation}
with discriminant $\Delta(k)$  explicitly given by

\begin{equation}
\begin{aligned}
    \Delta(k)=&\left[D_r+k^2D+\frac{K(k)}{2}\frac{\bar\rho}{D_c}\left(\mu\alpha-\frac{2}{3}\Omega\beta\right)\right]^2\\
    &-\left(k^2D+K(k)\frac{\bar{\rho}\mu\alpha}{D_c}\right)\left(2D_r+Dk^2-K(k)\frac{2\bar{\rho}\Omega\beta}{3D_c}\right)-\left(v-K(k)\frac{\bar{\rho}\mu\beta}{D_c}\right)\left(\frac{k^2v}{3}-K(k)\frac{2\bar{\rho}\Omega\alpha}{3D_c}\right).
\end{aligned}
\end{equation}

Note that the presence of the conserved field density $\drk$ implies that one of the eigenvalues must vanish at $k=0$. This property is satisfied by $\Lambda_2(k)$ and reveals an important feature of the unscreened ($\kappa=0$) case, i.e., in the limit $k\rightarrow 0$ and $\kappa\rightarrow 0$ do not commute. Indeed, if one takes the limit $\kappa\rightarrow 0$ before setting $k=0$, $K(k)=1$ becomes constant, inconsistently with $\Lambda_2(0)=0$. The first eigenvalue $\Lambda_1(k)$ is associated with a non-conserved mode that at $k=0$ reduces to $\Lambda_1(0)=-2D_r$, reflecting the fact that angular diffusion stabilizes the homogeneous disordered phase.
At small wave numbers and screened interactions $\kappa>0$, the eigenvalues behave as
\begin{equation}\label{eq:eigq0}
    \begin{aligned}
    \Lambda_1(k)&=-2D_r-k^2\left[\frac{\bar{\rho}\Omega}{3D_c\kappa^2}\left(-2\beta+v\alpha\right)-\frac{v^2}{6D_r}+D\right]+O(k^4),\\
    \Lambda_2(k)&=-k^2\left[\frac{\bar{\rho}\alpha}{D_c\kappa^2}\left(\mu-\frac{v\Omega}{3D_r}\right)+D_{\rm eff}(0)\right]+O(k^4).\\
    \end{aligned}
\end{equation}
It as apparent from equation \eqref{eq:eigq0} that for small values of $k$ the first eigenvalue $\Lambda_1(k)$ is stable due to the angular diffusion, while the eigenvalue $\Lambda_2(k)$ associated with the conserved field is stable for
\begin{equation}
   \frac{\bar\rho\alpha}{D_c}\left(\mu-\frac{v\Omega}{3D_r}\right) +\kappa^2 D_{\rm eff}(0)\ge 0.
\end{equation}

The last relation states that, at the macroscopic scale, the homogeneous phase is stable if, even in the case of effective attractive interaction among the particles, i.e., $\alpha(\mu-v\Omega/(3D_r))<0$, the traslational and orientational noise prevails. 
For fully unscreened interactions $k\gg\kappa\simeq 0$ and $K(k)\simeq 1$, at leading order in $\kappa/k$ the eigenvalues become 

\begin{small}
\begin{equation}
    \Lambda_{1,2}=-\left[D_r+\frac{\bar\rho}{2D_c}\left(\mu\alpha-\frac{2}{3}\Omega\beta\right)\pm\sqrt{\left[D_r+\frac{\bar\rho}{2D_c}\left(\mu\alpha-\frac{2}{3}\Omega\beta\right)\right]^2-\frac{\bar{\rho}\mu\alpha}{D_c}\left(2D_r-\frac{2\bar{\rho}\Omega\beta}{3D_c}\right)-\left(v-\frac{\bar{\rho}\mu\beta}{D_c}\right)\frac{2\bar{\rho}\Omega\alpha}{3D_c}}\right],
\end{equation}
\end{small}
which implies that, in the case of long-range interactions among the Janus colloids, at a large enough scale the system presents two non-conserved modes that can be both  stabilized by strong enough orientational disorder. Physically, being the number of particles a conserved quantity, in the limit $k\rightarrow 0$ there must always be a conserved mode that vanishes as $k^2$, signaling that the unscreened regime holds only for $\kappa\ll k$ and $k$ very small. 
Notably, the very same behavior has been found in the two species case in section \ref{sec:lambda34} for equal diffusivity and self-propelling velocity between the two species.

\section{ Effective interaction}\label{app:EffectiveInteraction}
Here we complement the calculations missing in section \ref{sec:effint}.
We are interested in solving the linear system in equation \eqref{eq:rholinear} with initial conditions $\delta\rho(\bm k,0)$ in the unstable regime. In particular, we need an expression for the ratio $\delta\rho_1(\bm k, t)/\delta\rho_2(\bm k, t)$ in the large $t$ limit.
This is captured by the ratio of the elements of the eigenvector corresponding to the eigenvalue $\Lambda_2$, associated with the most unstable mode which dominates the dynamics. The eigenvectors are given by

\begin{equation}
\hat{e}_{1,2}(k)=\left(
\begin{array}{c}
 \frac{D_1^{\rm eff}+\bar{\rho}_1 M_1 \Pi_1-D_2^{\rm eff}-\bar{\rho}_2 M_2 \Pi_2}{2
  }\pm\sqrt{\Delta} \\[5pt]
   \bar{\rho}_2 M_2 \Pi_1 \\
\end{array}
\right).
\end{equation}
In the non-oscillatory regime, the relative amplitude of the two density perturbations is given by

\begin{equation}
\begin{aligned}
     \frac{\delta\rho_1(\bm{k},t)}{\delta\rho_2(\bm{k},t)}
   &=\frac{\bar{\rho}_1 M_1}{\bar{\rho}_2 M_2}\frac{D_1^{\rm eff}+\bar{\rho}_1 M_1 \Pi_1-D_2^{\rm eff}-\bar{\rho}_2 M_2 \Pi_2-2\sqrt{\Delta}}{2\bar{\rho}_1 M_1\Pi_1},
\end{aligned}
\end{equation}
at leading order in $e^{k^2 \Lambda_2 t}$.

In the unstable oscillatory regime (blue area in figure \ref{fig:effectiveinteraction}) there is an alternation of depletion and aggregation in time according to

\begin{small}
\begin{equation}
\begin{aligned}
   &\frac{\delta\rho_1(\bm{k},t)}{\delta\rho_2(\bm{k},t)}=\\
   &\frac{\sin \left( \sqrt{|\Delta |}\, k^2 t\right) \left[\delta \rho_1(\bm{k},0) (-D_1^{\rm eff}+D_2^{\rm eff}-\bar{\rho}_1 M_1 \Pi_1+\bar{\rho}_2 M_2 \Pi_2)-\delta\rho_2(\bm{k},0)2
   \bar{\rho}_1 M_1 \Pi_2\right]+2\sqrt{|\Delta| }\, \delta \rho_1(\bm{k},0) \cos \left( \sqrt{|\Delta |}\, k^2 t\right)}{\sin
   \left( \sqrt{|\Delta |}\, k^2 t\right) \left[ \delta\rho_2(\bm{k},0)(D_1^{\rm eff}-D_2^{\rm eff}+\bar{\rho}_1 M_1 \Pi_1-\bar{\rho}_2 M_2 \Pi_2)- \delta \rho_1(\bm{k},0) 2\bar{\rho}_2 M_2  \Pi_1\right]+2\sqrt{|\Delta| }\, \delta\rho_2(\bm{k},0)\cos \left( \sqrt{|\Delta |}\, k^2 t\right)},
\end{aligned}
\end{equation}
\end{small}
where the common exponentially growing factor $e^{k^2\Trace{\mathcal{G}}/2}$ cancel in the ratio.
To better characterize this (linearly) oscillating phase we look at the rescaled variables $\delta\rho_{1,2}(\bm{k},t)e^{-k^2\Trace{\mathcal{G}}/2}$, whose time evolution describes an ellipse. The evolution of these two periodic trajectories can characterized by looking at their phase and amplitude as $\delta\rho_{1,2}(\bm{k},t)e^{-k^2\Trace{\mathcal{G}}/2}=A_{1,2}\cos(\sqrt{|\Delta|}k^2t+\varphi_{1,2})$, where

\begin{equation}
\begin{aligned}
   \tan\varphi_1 &= \frac{\delta \rho_1(\bm{k},0) (\mathcal{G}_{11}-\mathcal{G}_{22})+\delta\rho_2(\bm{k},0)2
   \mathcal{G}_{12}}{2\sqrt{|\Delta| }\, \delta \rho_1(\bm{k},0)},\\
   \tan\varphi_2 &= -\frac{\delta \rho_2(\bm{k},0) (\mathcal{G}_{11}-\mathcal{G}_{22})-\delta\rho_1(\bm{k},0)2
   \mathcal{G}_{21}}{2\sqrt{|\Delta| }\, \delta \rho_2(\bm{k},0)},\\
   A_1 
   &=2\sqrt{\mathcal{G}_{12}\left[ (\delta\rho_1(\bm{k},0))^2\mathcal{G}_{21}+\delta\rho_1(\bm{k},0)\delta\rho_2(\bm{k},0)(\mathcal{G}_{11}-\mathcal{G}_{22})+(\delta\rho_2(\bm{k},0))^2\mathcal{G}_{12}\right]},\\[5pt]
   A_2 
   &=2\sqrt{\mathcal{G}_{21}\left[ (\delta\rho_1(\bm{k},0))^2\mathcal{G}_{21}-\delta\rho_1(\bm{k},0)\delta\rho_2(\bm{k},0)(\mathcal{G}_{11}-\mathcal{G}_{22})+(\delta\rho_2(\bm{k},0))^2\mathcal{G}_{12}\right]},\\
\end{aligned}
\end{equation}
where we recall that complex eigenvalues exist only for $\bar{\rho}_1M_1\Pi_1<0$ and $\bar{\rho}_2M_2\Pi_2>0$.

\bibliographystyle{apsrev4-2}
\bibliography{references_NJP}

\end{document}